\documentclass[journal]{IEEEtran}

\usepackage{amsmath, amsthm, mathtools}
\usepackage{graphicx}
\usepackage{setspace}
\usepackage{amsfonts}
\usepackage{amssymb}
\usepackage{subcaption}
\usepackage{balance}
\usepackage{enumitem}
\usepackage{stfloats}
\usepackage{ltxtable,tabularx, tabu,longtable}
\usepackage{makecell}
\usepackage{multirow}
\usepackage[table]{xcolor}  
\usepackage{color}
\usepackage{pdfpages}
\usepackage{pdflscape}
\usepackage{everypage}
\usepackage{lipsum}
\usepackage{cite, url}
\newtheorem{mydef}{Definition}

\definecolor{mycolor}{rgb}{1,0.0,0.5}

\newcommand{\Lpagenumber}{\ifdim\textwidth=\linewidth\else\bgroup
  \dimendef\margin=0 
  \ifodd\value{page}\margin=\oddsidemargin
  \else\margin=\evensidemargin
  \fi
  \raisebox{\dimexpr-\topmargin-\headheight-\headsep-0.5\linewidth}[0pt][0pt]{%
    \rlap{\hspace{\dimexpr \margin+\textheight+\footskip}%
    \llap{\rotatebox{90}{\thepage}}}}%
\egroup\fi}
\AddEverypageHook{\Lpagenumber}%

\begin{document}

\title{On Coordination of Smart Grid and \\Cooperative Cloud Providers\thanks{\textcopyright 2020 IEEE Systems Journal.  Personal use of this material is permitted.  Permission from IEEE must be obtained for all other uses, in any current or future media, including reprinting/republishing this material for advertising or promotional purposes, creating new collective works, for resale or redistribution to servers or lists, or reuse of any copyrighted component of this work in other works.}}

\author{Monireh Mohebbi Moghaddam$^\ast$, Mohammad Hossein Manshaei$^\ast$, Mehdi Naderi Soorki$^\dagger$,\\ Walid Saad$^\ddagger$, Maziar Goudarzi$^+$, and Dusit Niyato$^\mp$\\
$^\ast$Department of Electrical and Computer Engineering, Isfahan University of Technology, Isfahan, Iran,\\
$^\dagger $ Engineering Department, Shahid Chamran University of Ahvaz, Ahvaz, Iran, \\
$^\ddagger$Department of Electrical and Computer Engineering, Virginia Tech, Blacksburg, VA, USA,\\
 $^+$Department of Computer Engineering, Sharif University of Technology, Tehran, Iran,\\
$^\mp$Department of Computer Science and Engineering, Nanyang Technological University, Singapore.}

\maketitle
\begin{abstract}
Cooperative cloud providers in the form of cloud federations can potentially reduce their energy costs by exploiting electricity price fluctuations across different locations. In this environment, on the one hand, the electricity price has a significant influence on the federations formed, and, thus, on the profit earned by the cloud providers, and on the other hand, the cloud cooperation has an inevitable impact on the performance of the smart grid. 
In this regard, the interaction between independent cloud providers and the smart grid is modeled as a two-stage Stackelberg game interleaved with a coalitional game in this paper. In this game, in the first stage the smart grid, as a leader chooses a proper electricity pricing mechanism to maximize its own profit. In the second stage, cloud providers cooperatively manage their workload to minimize their electricity costs. Given the dynamic of cloud providers in the federation formation process, an optimization model based on a constrained Markov decision process (CMDP) has been used by the smart grid to achieve the optimal policy. Numerical results show that the proposed solution yields around 28\% and 29\% profit improvement on average for the smart grid, and the cloud providers, respectively, compared to the noncooperative scheme. 
\end{abstract}
\section{Introduction}
\label{Sec:Intro}
The smart grid (SG), as the next generation of the power grid, which is characterized by two-way communications, advanced control technologies, and modern energy management techniques, enhances efficiency, sustainability, and reliability of the electric energy \cite{abbasinezhad2017TSG, abbasinezhad2019}.
One key feature of the SG is the so-called demand response (DR) model using which the SG can design suitable incentives to induce dynamic demand management of customers in response to grid conditions \cite{abid2019Electronics}. The utility companies can employ smart electricity pricing such as time or location-dependent pricing policies as DR programs to incentivize customers shift their load from peak to off-peak times or from one physical location to another \cite{el2016Allerton}.

Data centers (DCs) constitute a valuable target for DR due to their substantial energy consumption \cite{shehabi2018}. 
Moreover, loads of DCs are flexible that make them amenable to DR. DC operators can shift their load from peak to off-peak time, migrate their workload from one physical location to another,  or use the on-site generators and energy storage units. This flexibility provides an important opportunity for the SG to implement DR programs.
 Finally, the participation of DCs in DR programs is beneficial not only for the SG but also for the DC operators because it can reduce their electricity bill.  In this regard, cloud providers (CPs) have recently started investigating new approaches to develop suitable workload management mechanisms to increase their contributions in DR programs as well as to minimize their costs \cite{klingert2018ACM}. 
One promising solution for the power cost reduction of the cloud providers is through cloud cooperation in the form of \emph{cloud federations}. 
A cloud federation is a set of individual CPs agreeing to use their resources mutually and serve their users' workload cooperatively \cite{Guazzon2014CCGrid}.
Cloud federation enables the cloud participants to reduce their electricity bills by using more energy-efficient resources, accessing to more flexible energy management strategies, or by employing the spatial variation of the electricity prices over locations \cite{Giacobbe2015ComputerNetworks}. 

Particularly, when independent CPs cooperate with each other and shift their workload toward other DCs, their electricity demands change compared to the case of non-cooperation scheme. This not only changes the electricity bills of CPs but also has a major impact on the SG.  In practice, the SG seeks to balance supply and demand. In this regard, utility companies should predict their customers' demands and set electricity prices dynamically to achieve their goals. Geographic load distribution may lead to the changes in the power demands and it can also incur extra cost to the SG due to over/under loads if it is not considered by the SG \cite{chen2020Elsevier}.  
Despite this effect, the SG can utilize this opportunity by considering the possibility of cloud cooperation through the right setting the electricity prices for different power buses. Consequently, the CPs should decide on how to form the federations to minimize their electricity costs given the electricity prices. 

There have been a significant amount of efforts on designing effective programs to improve the participation of the DCs in DR (e.g., \cite{ bahrami2019CISS, chen2019TOMPECS}).  However, most of these works addressed the DR of one cloud provider and did not consider the interaction of multiple CPs with the SG. 
On the other hand, several contributions are focusing on the electricity bill reduction of DCs through geographical migration of the workload by forming the cloud federations (e.g. \cite{Guazzon2014CCGrid, Hassan2015}). However, these works did not consider the CPs cooperation in the context of demand response programs as well as the effect of this cooperation on the SG performance. 

Different from this prior literature, 
and regarding \cite{chen2020Elsevier}, which expressed the cooperation among multiple providers, and design of DR mechanisms for spatial-coupling loads, such as data centers as two main issues which require more attention from the research community, the main contribution of this paper is modeling and analyzing a situation in which an SG deals with independent CPs through a location-dependent DR program. To this end, we aim to formulate the process of dynamic location-dependent electricity pricing by the SG along with the cooperative workload management of independent CPs.
In particular, we consider an SG composed of multiple power buses and some DCs and model their interactions. 
The smart grid seeks to choose the set of prices to achieve an optimal balance between maximizing its profit and performing load balancing among power buses.
Meanwhile, the CPs will perform the demand side management by cooperatively servicing the users' workload by forming the federations to minimize their electricity costs.

For this scenario, we develop a game-theoretic model to jointly address the strategic decision of the SG for electricity prices and the cooperative workload management of CPs in response to the SG policies. This model is based on the controlled coalition game framework introduced in \cite{Niyato2011IEEE}. It is composed of a coalition formation scheme between the cloud providers and an optimization formulation for the smart grid. The proposed model is analogous to a nested two-stage Stackelberg game, integrated with a coalition formation game. The leading player in this game is the SG that chooses electricity prices for each bus, in the first stage. In the second stage, the CPs adapt their strategies (i.e. cooperative workload management through coalition formation) after observing the selected strategy by the smart grid. 
Our main contributions can be summarized as follows.

\begin{itemize}
\item \textit{Cooperative cloud providers and smart grid interaction}: We formulate the interaction between a smart grid and independent cloud providers who worked cooperatively in the form of cloud federations in the context of a dynamic electricity pricing scheme.

\item \textit{Modeling and solution techniques}: We model the problem as a two-stage Stackelberg game, integrated with a coalitional game. 
We apply the coalitional game to analyze the process of federation formation among CPs. Given the dynamic of  CPs, an optimization model based on the constrained Markov decision process (CMDP) used by the smart grid to find the optimal policy which is a mapping from coalition states to the electricity pricing policies.

\item \textit{Simulation and results}: We conduct simulations to show the effectiveness of the proposed approach. Numerical results show that the SG can achieve the objective of maximizing the revenue earned and minimizing the mismatch between the power supply and demand, and the energy cost of CPs reduces significantly.
\end{itemize}

The rest of this paper is organized as follows. In Section~\ref{Sec:RelatedWork}, we review related work. We introduce our system model in Section~\ref{Sec:SystemModel}. Section~\ref{Sec:problem_Formulation} gives the problem definition. 
In Section~\ref{Sec:Game_Formulation}, we present our game model. 
Simulation studies are presented in Section~\ref{Sec:Results}, and 
Section~\ref{Sec:Conclusion} gives the conclusion. 
\section{Related Works}
\label{Sec:RelatedWork}
We discuss two classes of related works.
The first category addresses the DR of geo-distributed DCs. The second category investigates techniques for exploiting cooperation among cloud providers for energy cost reduction. 
\subsection{Geo-distributed Data Center Demand Response}
\label{SubSec:DC_DR}
Although significant recent studies have investigated the problem of DC management in DR programs, only a handful of works looked at the participation of geo-distributed DCs. In this regard, the work in \cite{Zhou2015} and \cite{ zhou2018IEEE} designed an auction mechanism to elicit the potential of the demand response from a geo-distributed cloud.  \cite{yu2016TCC} proposed a  price-sensitivity aware geographical load balancing scheme for distributed Internet data centers taking into account the dynamic characteristics and actual physical constraints of the SG. The authors in \cite{gu2018Elsevier} focused on the energy cost and carbon emission of cloud DCs equipped with renewable energies, and study the green scheduling of them by energy trading with the SG. 

DR of distributed DCs in the deregulated energy market considering the uncertainties in the arrival rates of the workloads, local renewable generation, and time-varying electricity prices is studied in \cite{Bahrami2017IEEE} and \cite{ bahrami2019CISS}.  
In \cite{Zhou2015ACM}, authors studied the pricing of bilateral electricity trade between a cloud with geo-distributed hybrid DCs and the corresponding smart grids. 
Moreover, in \cite{Wang2014PES} and \cite{Tran2016}, authors formulated the participation of geo-distributed DCs in a dynamic electricity pricing program as a two-stage Stackelberg game. In these prior works, the smart grid is modeled as a leader that performs load balancing among power buses and maximizes its profit, whereas the cloud provider as the follower seeks to maximize its benefit.       
The authors in \cite{Wang2016IEEE} studied the same problem while taking into account the active decisions of both sides and formulated the problem using bilevel programming. A summary of the recent researches on the participation of geo-distributed DCs in DR programs is shown in Table \ref{Table:RelatedWorks1}.

In contrast to this prior art, we address the problem of demand response of independent CPs, not a geo-distributed cloud.
 In the case of geo-distributed DCs belonging to a cloud, all DCs always collaborate with each other to maximize the profit of the corresponding cloud. 
This assumption is not true for independent CPs. As they are often selfish entities who cooperate with each other only when
it is beneficial for them. Therefore, when dealing with the DR of multiple CPs, one must investigate what stable coalitions are formed and how the smart grid affects this collaboration.
\subsection{Energy Management of Cloud Federation}
\label{SubSec:Cloud_Federation}
There is a significant number of works that addressed the problem of federation formation among CPs for various targets such as improving resource utilization or overcoming the resource limitation (e.g. \cite{ElZan2014CloudNet, Hassan2014, Bairagi2016ICOIN, Niyato2011ICST}).
 However, only a few recent works focused on the federation formation for the purpose of energy cost reduction \cite{Hassan2015, Kecskemeti2013, Guazzon2014CCGrid, Kaewpuang2014IEEE} as well as energy sustainability \cite{Celesti2013CLOSER}. 

Authors in \cite{Hassan2015} proposed an energy-aware resource and revenue sharing mechanism and focused on minimizing the overall energy of cloud providers.
The work in \cite{Guazzon2014CCGrid} addressed the problem of federation formation in order to reduce the CPs' energy cost and devised an algorithm based on cooperative game theory to allocate workload in an energy-aware fashion. 
In \cite{Kaewpuang2014IEEE}, the authors proposed a framework for cooperative virtual machine (VM) management of cloud users in a smart grid environment. They designed an algorithm to allocate VM of cloud users to the available resources and to manage the DCs' power consumption under various uncertainties.
 
The prior works in \cite{Hassan2015, Kecskemeti2013, Guazzon2014CCGrid, Kaewpuang2014IEEE} assumed that the electricity prices are not affected by the CPs' cooperation. Consequently, they did not consider the impact of the cloud federation on the smart grid as well as how the smart grid takes its decision while taking into account the cooperative workload management of the CPs. Unlike these studies, we consider both the impact of the smart grid decision on the cooperative VM allocation among the cloud providers and the influence of the cloud providers' cooperation on the smart grid. 
  
 Beyond these categories, \cite{Niu2016ACM}, \cite{yu2017ACM} and \cite{mohebbi2019IEEE} are three relevant studies  which addressed DR of cooperative CPs. The authors in \cite{Niu2016ACM} presented the aggregated participation of the CPs in a smart grid's power reduction program. 
 The work in \cite{yu2017ACM} modeled the  DCs electricity procurement process in the wholesale electricity market. In this work, DCs aggregate their loads to mitigate their power demand uncertainty in the market.
Leveraging the cooperative game theory,  \cite{mohebbi2019IEEE} proposed a cloud federation formation algorithm in the presence of a location-dependent DR program and analyzed the effect of this cooperation on the SG performance.
In contrast to the works in \cite{Niu2016ACM} and \cite{yu2017ACM}, we consider the cooperative workload management of geo-distributed cloud providers. Because of the geographical distribution of CPs, they can take advantage of the variety in the electricity prices to reduce their costs. Besides, unlike these works which consider the aggregation of CPs demands, in our work, VMs are migrated from one CP toward the others to reduce the electricity bill. 
Compared to \cite{mohebbi2019IEEE} in which the SG is unaware of the CPs cooperation, here, we consider the SG as an active agent in our game model, takes strategic decision and model the interaction of the SG with CPs along with the CPs cooperation. 
The related works in this category are summarized in Table \ref{Table:RelatedWorks2}.
 \section{System Model}
 \label{Sec:SystemModel}
We consider a system that consists of multiple cloud providers with geographically distributed data centers that interact with a smart grid system. 
Each data center is connected to a power bus to obtains its power.
The smart grid operator should denote, and consequently announce the price function of each bus. Given pricing functions, each CP should specify the amount of the electricity demand of its data center. The power demand of each DC is denoted by the amount of the workload that should be processed. 
 Each cloud provider can individually process its users' requests or can cooperate with the others if the cooperation leads to more benefits compared to the noncooperative case. In the following, we describe our system model with more details.
\subsection{Smart Grid}
\label{SubSec:SGModel}
We consider a discrete time model $ \mathcal{T }=\{1,..,T\} $ in which the length of each time slot $ t $ is determined according to the time-scale at which the pricing decisions are updated (for e.g. one hour).
We consider a smart grid that encompasses a set $\mathcal{I}$ of $ I $ power buses.
The power buses are interconnected through branches and form grid topology. Each power bus is connected to various load devices. In our model, some buses include data centers that provide cloud computing services \cite{Wang2014PES, Wang2014IEEE}. In essence, we consider a set $\mathcal{N}$ of $ N $ distributed DCs belonging to different cloud providers. Each data center $ j $ is connected to one power bus in the smart grid to obtain its required electricity \cite{Niu2016ACM, yu2017ACM, mohebbi2019IEEE}.
As we are interested in demand response of data centers, we focus on these loads and assume that non-data center power demands
are not elastic and the smart grid can accurately predict their demands for the near future \cite{Wang2014PES, Wang2014IEEE}.  
 
 As the DR program, we consider a location-dependent pricing scenario such as \cite{Wang2016IEEE, Wang2014IEEE, Wang2016IGSEC} for charging data centers. In this regard, we define the electricity price function $ \theta_j^t$ for DC $ j $ at time slot $ t $ as follows:
\begin{equation}
\label{Eq:ElecPrice2}
\theta_j^t=\beta_j(e_{j}^t-\delta_{j}^t)+z_{j}^t,
\end{equation}
where $ e_{j}^t $ represents the  power consumption of data center $ j$, $ \delta_j^t$ is the billing reference should be determined by the smart grid, $ \beta_j >0 $ is a sensitivity parameter, and $ Z_{j} ^t$  is the base electricity price of the power bus $j$ at time $ t $. This dynamic pricing scheme motivated by the tiered electricity pricing, which has been widely implemented in many power markets. 

To encourage the CPs to shift their electricity loads to less loaded locations and prevent the SG from abusing its market power, some constraints should be set to regulate the electricity prices. In practice, the smart grid and cloud providers usually negotiate with each other and enter into a contract to specify the pricing structure. Based on related studies such as \cite{Wang2016IEEE}, electricity prices should set in a specific range $ [\theta_{l} , \theta_{h}] $, where $  \theta_{l} $  and  $ \theta_{h} $ are lower and upper bounds, respectively.
  \subsection {Cloud Computing}
\label{SubSec:CCModel}
In our system, each data center $ j\in \mathcal{N} $ consists of a
set  $\mathcal{M}_j $ of ${M}_{j} $ hosts. For simplicity, we assume that all of the hosts that belong to one of the DCs are homogeneous in terms of processing capacity and provided the amount of memory \cite{Wang2016IGSEC}. CPs provide their physical resources in the form of virtual machine (VM) instances to customers. Each  CP can offer different VM classes which are different in terms of the processing capacity as well as the available memory.  Without loss of generality, we assume that all CPs offer the same VM class.

A VM allocated on a physical host uses a certain fraction of processing capacity and a specific amount of the memory. Generally, each host has adequate resources for running more than one VM at a specific time. Let $a_{j} $ be the total number of VMs that can be served simultaneously by a single host of DC $ j $. In this regard, we define $ Q_j$ as the processing capacity of CP $ j $, which is equal to $M_ja_j $ \cite{Guazzon2014CCGrid}.

 In a specific time slot, the amount of the workload of users of CP $ j $ is represented by $ \mathcal{W}_j$, which is the set of VMs that compose this workload. We denote the number of requested VMs of CP $ j $ at time slot $ t $ by $ W_j^t$. CP $ j$ can process this workload either by using its hosts or by shifting it to the DCs of other CPs. Additionally, each CP $ j$ defines a revenue policy for charging its customers. In other words, CP $ j $ should determine a per unit price for each VM class that customers should pay. 
 In this way, we define $ r_j$ as the CPs charge for processing each VM from class $ j $ \cite{Guazzon2014CCGrid}.
\subsection{Data Center Power Consumption}
\label{SubSec:DC_Power}
Here, our focus is on the energy used by hosts while ignoring the energy consumption of the other facilities such as cooling devices and unit power supplies. However, the power consumption of these facilities is roughly proportional to the host power \cite{Qureshi2010} through the power usage efficiency (PUE). PUE is defined as a ratio between the total power amount used by the entire DC facility (consisting of hosts, cooling devices,  etc.) and the power consumption of IT equipment. 

We model the host power consumption as $P^\textrm{idle}_{j}+U (P^\textrm{peak}_{j}-P^\textrm{idle}_{j})$ for any host $ j $. 
Here, $ P^\textrm{peak}$ denotes the host power when it is fully utilized, $ P^\textrm{idle}$ represents the amount of power consumption in an idle state, and $ U$ is the fraction of CPU being used. In \cite{Borgetto2012Elsevier}, the authors showed that this model, despite the simplicity, provides an accurate estimate of the power consumption for different host types. 
Hence, when $ m_j^t $ hosts are active in time slot $ t $, the average power consumption of DC $j$  can be calculated as follows \cite{Niu2016ACM}:
\begin{equation}
\label{Eq:PowerEquation}
 e_{j}^t=m_j^t\left[P^\textrm{idle}_{j} +\left(P^\textrm{peak}_{j}-P^\textrm{idle}_{j}\right)U_j^t\right]   \gamma_j,
\end{equation} 
where $ U_j^t$ is the average CPU utilization level of hosts at DC $j$, and $ \gamma_j $ is its PUE. This parameter usually has a value in the range of $[1.1,3]$. A smaller value indicates a more efficient DC in terms of power consumption. Table \ref{Tab:Notations} lists all of used notations.
 \begin{table}[t]
 \centering
\rowcolors{2}{gray!20}{white}
 \caption{List of Notations.}
\begin{tabular}{c|l}
 \hline
 Notation& Definition \\ \hline 	\hline
$ \mathcal{I}$& Set of power buses\\
$ \mathcal{N}$& Set of CPs \\  
$ \mathcal{T}$& Set of time slots\\  \hline
 $ \theta_j^T  $&  Unit electricity price for DC $ j $ at time slot $ t $\\ 
$ z_j^t $& Base electricity price at power bus $ j $ at time slot $ t $\\ 
   $  \beta_j $ & Sensitivity parameter at price function\\
 $ \delta_j^t $ & Billing reference at bus $ j $ at time slot $ t $\\
 \hline
$ M_j$& Number of hosts of DC $ j $\\ 
 $ W_j^t$& Workload of CP $ j $ at time slot $ t $\\ 
 $ Q_j$& Processing capacity of DC $ j $\\ 
  $ r_j$& Charging price for a VM of CP $ j $\\ 
 $ a_j$& Number of VMs served by a single server\\   \hline
 $ e_j^t $& The amount of power consumption of  DC $ j$\\ 
 $  P_j^\textrm{idle} $& Average idle power of a host at DC $ j $\\ 
 $  P_j^\textrm{peak} $& Average peak power of a host at DC $ j $ \\
 $ m_j^t $ & The number of active hosts of  DC $ j $ at time slot $ t $\\ 
 $ \gamma_j $& Power usage efficiency (PUE) of DC $ j $\\ 
 $ U_j^t $& Average CPU utilization of servers at DC $ j $ \\ \hline
 $\mathcal{S}$& A CP coalition/federation \\ 
 $ v(\mathcal{S}) $& Worth of coalition $ \mathcal{S} $ \\ 
 $ \theta_\mathcal{S} $& Electricity prices for CPs in coalition $ \mathcal{S} $\\ 
 $ \psi_j(\mathcal{S}) $& Payoff received by CP $ j $ as a member of $ \mathcal{S} $\\  
  $ R(\mathcal{S}) $& Revenue rate of coalition $ \mathcal{S} $ \\ 
 $ C(\mathcal{S}) $& Cost rate of coalition $ \mathcal{S} $ \\ 
 $ \omega_{i,j}^t$& Number of migrated VMs from CP $ i$ to $ j$ \\ 
 $ C_{i,j}^{M} $& Cost of VM migration from CP $ i$ to $ j$ \\ \hline
$ U^{SG}  $& The SG utility function\\ 
$ G_j^t $& Maximum available power supply to DC $ j $\\ 
$ \theta_l  $&  Lower bound on the electricity price \\ 
$ \theta_h  $&  Upper bound on the electricity price \\ 
$ \alpha_1,\alpha_2 $& Weighting parameters in the SG utility function\\  \hline
$ \varOmega $& State space of CPs coalition formation game \\ 
 $ \chi $& A coalition partition/state \\ 
$ B_N $& $ N $-th Bell Number \\ 
 $  \boldsymbol{T}(\boldsymbol{\delta}) $& Transition probability matrix of a coalition state \\ 
  $ \eta_{k,k^{'}}(\boldsymbol{\delta}) $& Prob. of changing state from partition $ \chi_{k} $ to $ \chi_{k^{'}} $ \\ 
  $ \sigma $& Prob. of performing merge and split  by a CP\\ 
  $ \tau(.) $& Best-reply rule function \\ 
 $  \varrho $& The probability of a CP acting rationally \\ 
  $ \epsilon $& The probability of a CP acting irrationally \\ 
  $ \overrightarrow{\boldsymbol{p}}$& Stationary probabilities of Markov chain\\
$ \pi $& The SG pricing policy \\ \hline
\end{tabular}
 \label{Tab:Notations}
 \end{table}
  \section{Problem Statement}
 \label{Sec:problem_Formulation}
\begin{figure}[t]
    \centering
    \includegraphics[scale=0.50]{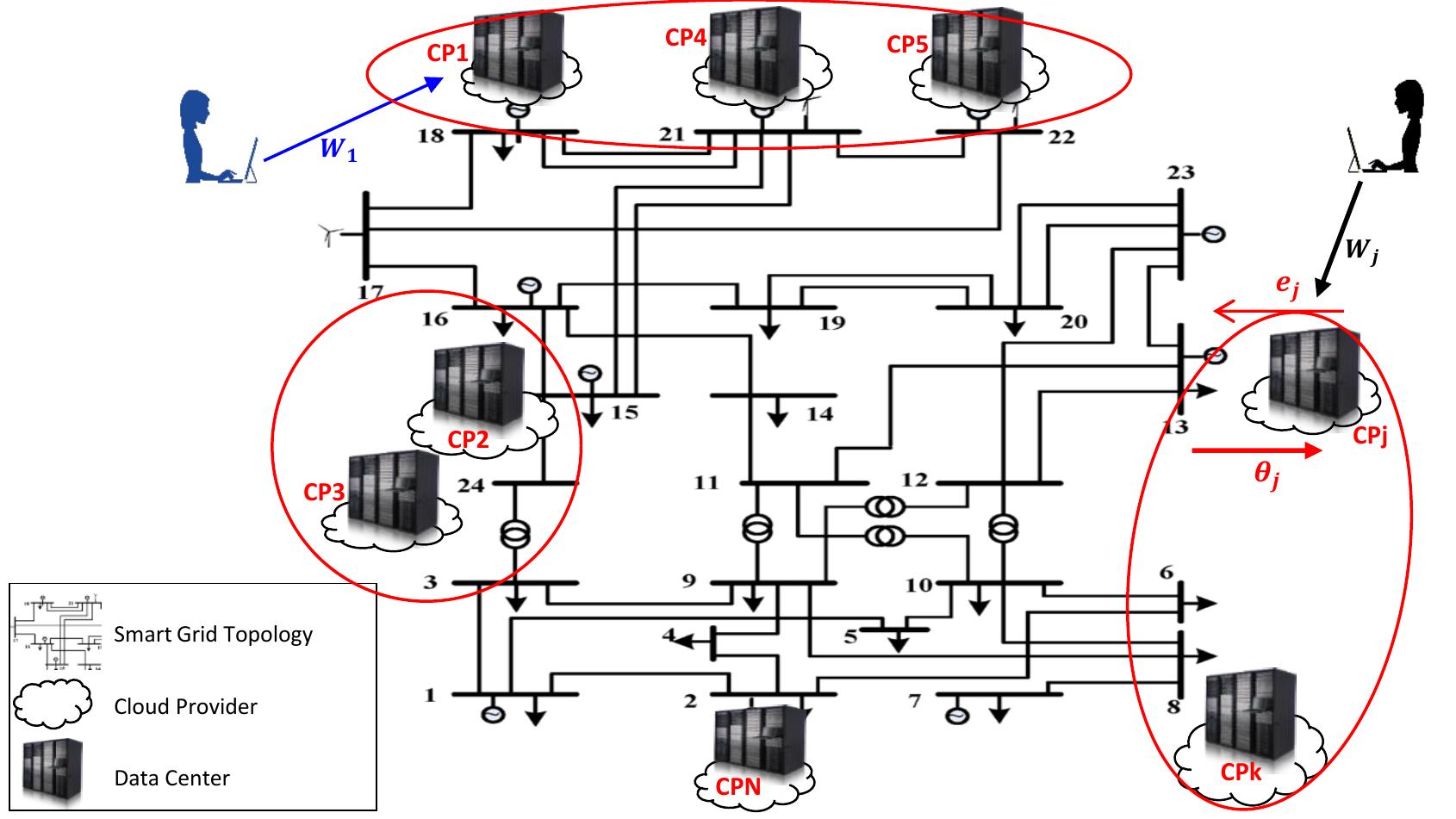}
    \caption{An overview of the system model and problem statement.}
    \label{Fig:System_Model}
    \end{figure}

Given the defined system model, we can now describe our problem through an example. 
Consider the smart grid system in Figure~\ref{Fig:System_Model} which represents the IEEE 24-bus reliability test system, which includes 24 power buses and 38 branches. Assume that there are $ N $ data centers in the system connected to different buses. Each data center is in the domain of a specific cloud provider, and obtains its power through the bus is connected to.
In this system, as stated before, the SG should announce the price function for each power bus, according to (\ref{Eq:ElecPrice2}).  In this regard, as $ \beta_j $ and $ z_j^t$ parameters have known and predictable values, the SG should determine the value of $ \delta_j^t $. The smart grid seeks to choose the set of prices to achieve its predetermined goals which are maximizing the revenue and minimizing the mismatch between the supply and demand. Given electricity pricing functions, each CP should specify the power demand of its data center, i.e. $ e_j^t$ value in  (\ref{Eq:ElecPrice2}). 
Here, CPs are flexible, rational customers who perform the demand side management in such a way maximize their profits. The profit defined as the total revenue obtained from serving the users' requests, subtracted by the energy cost paid to the SG.

 As shown in this figure, each data center $ j$ receives some requests for workload processing from its users, denoted by ${W}_{j} $. In a noncooperative case, each cloud provider should process its own workload and, then, submits the power demand to the smart grid in a way to ensure a service for all of its workload. However, cloud providers can cooperate with each other in the form of cloud federations and jointly process their workload in such a way to reduce their electricity cost. For example, consider a case with N CPs in which CPs $ 1 $, $ 4 $ and $ 5 $ form a coalition with three members, CPs $ 2 $ and $ 3 $ form a coalition, CPs $ j $ and $ k $ form another coalition, and other CPs prefer to work alone. In this case, the users' requests received by a coalition member can migrate to the other one, if this migration leads to less cost than working non-cooperatively. 

Within the aforementioned system, on the one hand, the coalitions formed among CPs depend on the pricing functions denoted by the SG. On the other hand, the electricity prices as well as the profit earned by the SG influenced by the coalition formation process performed by CPs.
 In other words, the CPs form the coalitions in such a way minimize their energy cost and changing the electricity price functions affect on their cost. Hence, the SG can affect on the CPs' decisions and profits by controlling the pricing functions. Indeed, the coalition formation among CPs changes their power demand of different buses, consequently, has an undeniable influence on the SG's profit. 

Given the above description, it is clear that the selected strategy by each side (i.e., the SG and CPs) strongly affects the behavior and subsequently, the profit of the other side. Therefore, we are facing a strategic situation with rational agents where the action taken by each decision-maker influences on the behavior and the payoff of the other. 
 In such setting, game theory is a proper analytical tool that can be utilized to formulate the interactions between the SG and CPs from one side and the cooperation among CPs from the other side \cite{Niyato2011IEEE}. In the next section, we will describe a game model that is designated to specify the optimal strategies should be taken by decision makers.
\section{Game Model}
\label{Sec:Game_Formulation}
\begin{figure*}[t]
  \centering
   \includegraphics[scale=0.95]{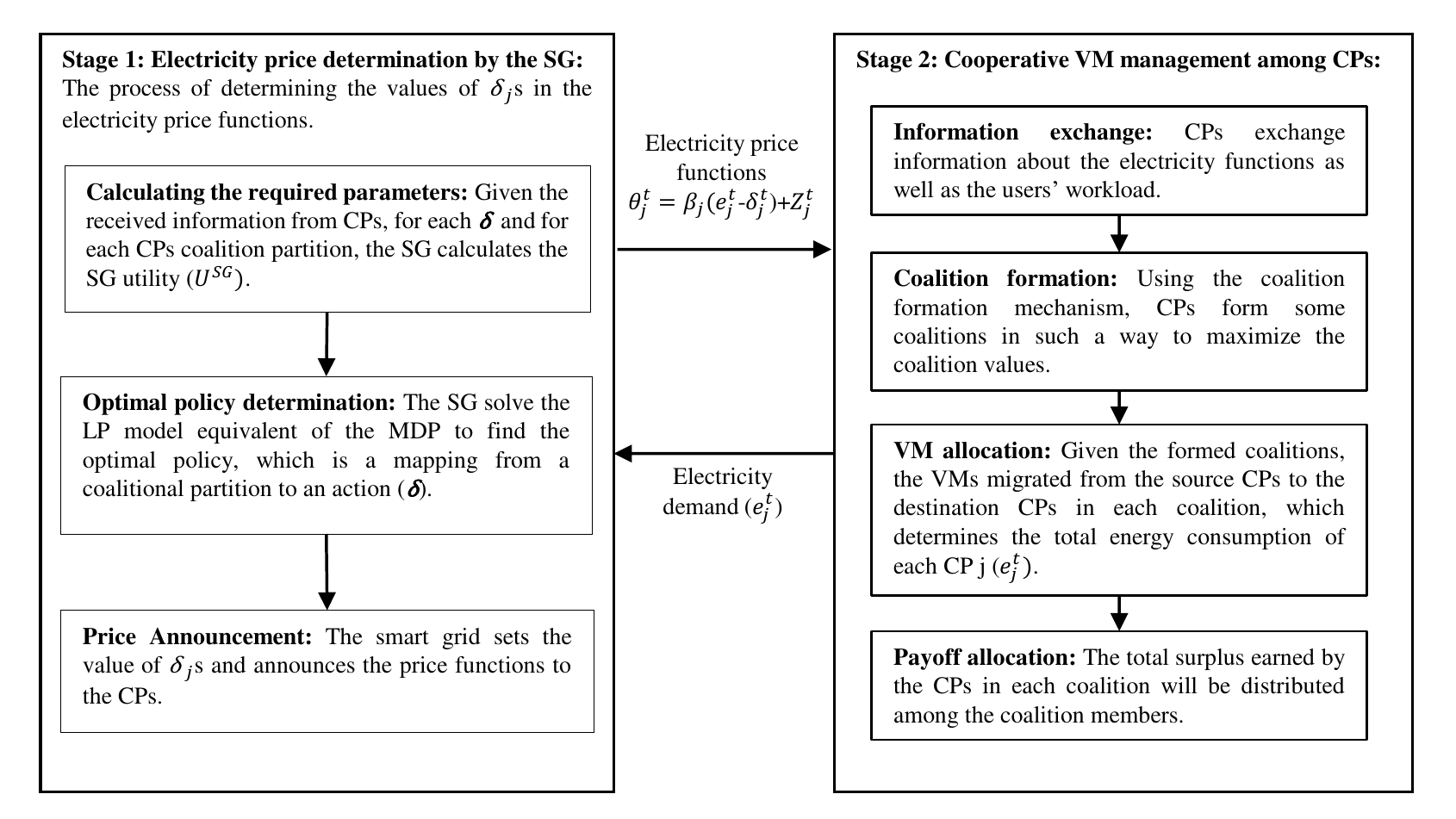}
    \caption{Hierarchical two-stage game model. In the first stage,
  the SG chooses its strategy (the electricity price functions) so that maximize its profit. In stage 2, the CPs adapt their strategies (i.e. coalition formation) to minimize their energy cost.}
  \label{Fig:Hirrarchical_Model}
  \end{figure*}

In this section, we present our game model to jointly address the coalition formation between the CPs and finding the optimal pricing strategy of the SG. We need a hierarchical model in which the SG sets the pricing functions at the first and, then, the CPs respond to the SG action by forming proper coalitions. In this regard, we designate a two-stage game model, integrated with a coalition formation game in the second stage, named "Interactive Cooperative Game (ICG)". 
This model which is based on the controlled coalition game introduced in \cite{Niyato2011IEEE}, represented in Figure~\ref{Fig:Hirrarchical_Model}, and its details will be explained in the following subsections. Particularly, in this model, the SG as the leading player should specify the pricing functions given the possible formed coalitions among CPs, such that maximize its own profit. 
Furthermore, in the second stage, the CPs as the followers manage their workload collaboratively so that they minimize their cost.

To obtain the optimal strategies for the leader and followers, our game is decomposed into the following two interrelated problems: 1) An optimization formulation for the SG and 2) Cooperative VM management by forming the CPs coalitions in a distributed way. We start with the problem of cooperative workload management by the CPs, and, then formulate the optimization formulation for the SG. 
\subsection{CPs Coalitional Game}
\label{SubSec:Coalitional_Game_Model}
In this section, we study how CPs form stable coalitions,
 optimally allocate the coalition workload and divide the coalition profit. We apply the coalitional game theory to analyze the cooperative behavior of CPs. Coalitional game is a branch of game theory focused on studying the behavior of a group of rational agents when they cooperate \cite{myerson2013, han2019}. 
 In particular, coalition formation game \cite{myerson2013} is a special class of coalitional games used for analyzing the coalition
formation process between the players. In our model, we should
investigate the cooperative workload management of CPs by forming the federations and analyzing their characteristics. Hence, this type of cooperative games is appropriate to use.
\subsubsection{Coalitional Game Characterization}	 
\label{SubSec:Coalition_Game_Characterization}
Given the set of CPs as players of the coalition game,
a coalition $\mathcal{S} \subseteq \mathcal{N}$ is defined as a non-empty subset of CPs who agree to cooperatively serve their workload. 
We denote a coalition structure/partition of $ \mathcal{N}$ by $ \chi $, where $\chi=\{\mathcal{S}_1,\mathcal{S}_2,...,\mathcal{S}_l \} $. In our coalitional game model, each provider is a member of exactly one coalition, i.e., $ \mathcal{S}_i\cap
\mathcal{S}_j=\varnothing, \forall i,j $, where $ i \neq j $ and
$\bigcup_{i=1}^{l} \mathcal{S}_i = \mathcal{N} $. The
total number of coalition structures is $ B_N $, where $ B_N $, the $ N $-th Bell number, is the number of nonempty subsets a set of size $ N $ can be partitioned into \cite{Knuth2011}. 

In addition to the player set, each coalition $ \mathcal{S} $ has a coalition value, denoted by $ v(\mathcal{S}) $.
Mainly, the coalition value quantifies the net profit of coalition
$\mathcal{S}$. In the following, we define the coalition value for our coalition formation game.
\subsubsection{Coalition Value}
\label{Coalition_Value_Calculation}
To determine the coalition value, we should define the revenue and cost rate of a coalition $ \mathcal{S}$, which are represented by $R(\mathcal{S}) $ and $ C(\mathcal{S}) $, respectively. 
 $R(\mathcal{S}) $ is the sum of revenue rates of individual VMs which served by the coalition members, and can be calculated as: 
 \begin{equation}
 \label{Eq:Coalition_Revenue}
R(\mathcal{S})= \sum_{j \in \mathcal{S}} {W}_j^tr_{j}.
\end{equation}
The associated cost consists of the energy cost of processing the coalition workload, which depends on the values of $ \theta_j^t$, plus the cost of the VM migration between the coalition members (e.g., due to network bandwidth used for migration). This is defined as:
  \begin{equation}
 \label{Eq:Coalition_Cost}
 C(\mathcal{S}) = \sum_{j\in \mathcal{S}} \theta_j^t(\boldsymbol{\delta}) e_j^t(\boldsymbol{\delta})-\sum_{i \in \mathcal{S}}  \sum_{j \in \mathcal{S}}  \omega_{i,j}^t C_{i,j}^{M},
 \end{equation}
  where $  C_{i,j}^{M} $ is  the hourly migration cost of a VM from CP $i  $ to CP $j $, and $ \omega_{i,j}^t$ represents the number of migrated VMs. $ \theta_j^t$ is the electricity  price at bus $ j$, and obtained from (\ref{Eq:ElecPrice2}). $ e_j^t $ is the amount of the power consumption of CP $ j $, and computed from  \ref{Eq:ElecPrice2}.
 The VM migration will be performed when the total energy cost of processing a VM in the destination DC, plus the migration cost becomes less than the cost of serving that VM in the source DC. This difference generally stems from the location-dependent electricity prices. Given the above explanation, we define the coalition value as follows:
\begin{equation}
 \label{Eq:Coalition_Val}
 v(\mathcal{S})=R(\mathcal{S}) - C(\mathcal{S}).
\end{equation}

As we can see from (\ref{Eq:Coalition_Revenue})-(\ref{Eq:Coalition_Val}), 
the coalition value $ v(\mathcal{S}) $ solely depends on the members of $ \mathcal{S} $, with no dependency on the other coalitions outside $\mathcal{S}$.
This property indicates the class of cooperative games with characteristic form. So, our game belongs to this class.  Moreover, the proposed game has a transferable utility (TU) because the coalition value in (\ref{Eq:Coalition_Val}) is the amount of money gained by this coalition and can be divided among the members in any arbitrary manner \cite{Saad2009IEEE}.

We assume that all CPs apply the same policy to charge their customers (same values for $ r_j $ in  (\ref{Eq:Coalition_Revenue})). So, the
associated cost of a coalition $ \mathcal{S} $, i.e., $ C(\mathcal{S}) $ in (\ref{Eq:Coalition_Val}), plays an important rule in the coalition value. 
This cost depends on how the coalition workload is assigned to the hosts of the coalition members. As a result, different VM allocations change the values of $ e_j^t $ and $ \omega_{i,j}^t $, subsequently, $ v(\mathcal{S}) $.
Moreover, as CPs are rational players, they seek to minimize their cost as much as possible. Therefore, we should find the optimal solution of the VM allocation for a given coalition $ \mathcal{S} $, so as to minimize the total cost. We cast this problem as a VM allocation optimization.
\subsubsection{Optimal VM Allocation}
\label{VM_Allocation}
As we mentioned, the VM allocation should be done in such a way the cost becomes minimized. Hence, we can formulate the optimal VM allocation as follows:
\begingroup
\allowdisplaybreaks
\begin{align}
&\min_{\omega_{i,j}^t} \qquad  C(\mathcal{S}) \label{Eq:CP_opt}\\
\textrm{s.t.} \qquad 
& \sum_{j \in \mathcal{S}}{ \omega_{i,j}^t } = W_i, ~\forall i \in \mathcal{S},~\forall t \in \mathcal{T},\label{Eq:CP_opt_c1} \\
& \sum_{i \in \mathcal{S}}{\omega_{i,j}^t} \leq Q_j ,~\forall j \in \mathcal{S},~\forall t \in \mathcal{T}, \label{Eq:CP_opt_c2}\\
& \omega_{i,j}^t \in \mathbb{Z}^{+},  ~\forall i, j \in \mathcal{S}, ~\forall t \in \mathcal{T}, \label{Eq:CP_opt_c3} 
\end{align}
\endgroup
where $ \omega_{i,j}^t$ are decision variables, which are the number of migrated VMs from DC $ i $ to $ j $. The value of  $ \omega_{i,j}^t$ affects on the number of active hosts. (\ref{Eq:CP_opt_c1}) ensures that all workloads are served. The condition presented by (\ref{Eq:CP_opt_c2}) guarantees that the total workload assigned to a CP will not exceed its capacity. The final constraint  (\ref{Eq:CP_opt_c3}) defines the valid domain of the decision variables.
By solving the optimization problem,  the workload should be migrated ($ \omega_{i,j}^t$) and processed by each CP will be specified. As a result, the amount of energy consumption of each data center ($ e_j^t $) is determined.
\subsubsection{Payoff Allocation}
\label{Payoff_Allocation}
The final output of the optimization problem in  (\ref{Eq:Coalition_Cost}) is the optimal VM allocation among the coalition members. In this stage, the coalition value, $ v(\mathcal{S}) $, can be calculated from  (\ref{Eq:Coalition_Val}), by substituting the coalition cost from the revenue earned. Now, we should determine how the total surplus should be distributed among the coalition members. The payoff allocation rule specifies this allocation. In this regard, we apply the Shapley value \cite{Shapley1988} to divide the coalition payoff fairly. Here, fairness indicates that CP's profit is proportional to the value it is added to the federation. In other words, the Shapley value represents the marginal contributions of any CP to the federation it belongs to. In contrast with other payoff sharing rules, this scheme allows federations to allocate payoff to their members according to their economical contributions. As the CPs seek to maximize their monetary payoffs and cooperate with the others when the cooperation leads to more benefit from acting alone, this scheme is a good choice.

The Shapley value associates with every coalition a unique payoff vector. This value for $ j \in \mathcal{S} $ in a coalition formation game with TU
is calculated as follows:
 \begin{equation}
 \label{Eq:Shapley_Value}
  \psi_j(\mathcal{S})=\sum_{\mathcal{F} \subseteq \mathcal{S}\{j\} } \dfrac{|\mathcal{F}|!(|S|-|\mathcal{F}|-1)!}{|S|!}(v(\mathcal{F}\cup {j})-v(\mathcal{F})),
 \end{equation}
 where $ \mathcal{F} $ denotes all subsets of $ \mathcal{S} $ that do not include $j $.  In other words, $ \psi_j(\mathcal{S}) $ determines the final payoff that received by cloud provider $  j$, from the coalition $ \mathcal{S} $.  
It is clear from Equations (\ref{Eq:Coalition_Cost})- (\ref{Eq:Shapley_Value}) that the electricity prices have a direct effect on the profit earned by the CPs.
\subsection{The Smart Grid Utility}	 	 
\label{SubSec:SG_Utility_CentScheme}
Before introducing the interactive game model,  we should specify the SG's utility function. 
We define the objective of the SG as an optimal balance between maximizing its revenue earned from selling the power and minimizing the mismatch between the power supply and demand. 
In this regard, we consider a utility function similar to \cite{Wang2014PES} as follows:
   \begin{equation}
   \label{Eq:SG_Utility}
    U^{SG}(\chi_k,\boldsymbol{\delta})=\alpha_1 \sum_{j=1}^{N}{\theta_j^t(\boldsymbol{\delta}) e_{j}^t(\boldsymbol{\delta})}\\
    -\alpha_2K \sum_{j=1}^{N}{|e_{j}^t(\boldsymbol{\delta})-G_{j}^t|},
     \end{equation}
 where $ \chi_k$ is a partition of CPs, $ K $ is a normalized parameter, and $ G_j^t$ is the maximum available power supply to DC $ j $ at time M, which is difference between the total power capacity at that bus and the background load. The first term in  (\ref{Eq:SG_Utility}), specifies the total smart grid's revenue from selling electricity to the CPs, and the second part measures the amount of imbalance between the power generation and  load. In (\ref{Eq:SG_Utility}), $ \alpha_1$ and $ \alpha_2$ are some wights, determined according to the importance of each term, where $ \alpha_1+\alpha_2=0$, $ 0\leq \alpha_1,\alpha_2\leq1$ . It is clear from (\ref{Eq:SG_Utility}) that the utility of SG changes by the electricity prices, or the SG strategy ( $ \boldsymbol{\delta} $) as well as the CPs' power demand ($\boldsymbol{e} $). As we mentioned in Section \ref{SubSec:Coalitional_Game_Model}, the power demand of CPs depend on the partition formed among them ($ \chi_k$). Hence, each pair $ (\chi_k,\boldsymbol {\delta}) $ leads to a specific profit for the SG.
 
The SG would like to directly control the CPs coalition formation process and determine the final CPs' partition in a centralized manner, if the CPs fully cooperate with SG. In this way, the SG operator selects the partition and price set to maximize its payoff. Hence, we have the following optimization problem in this scheme:
 \begingroup
 \allowdisplaybreaks
 \begin{align}
 &\max_{\chi_k,\boldsymbol{\delta}} \qquad  U^{SG}(\chi_k,\boldsymbol{\delta}) \label{Eq:SG_opt}\\
 \textrm{s.t.} \qquad 
 & \theta_{l} \leq \theta_j^t  \leq \theta_{h}, \forall j \in \mathcal{N},\\
 & \mathcal{S} \cap \mathcal{S'}=\emptyset, \forall \mathcal{S},\mathcal{S'} \in \chi_k,\\
 & \cup_{\forall \mathcal{S} \in \chi_k }\mathcal{S}=\mathcal{N}.
 \label{Eq:SG_opt_cons.} 
 \end{align}
 \endgroup
 where $ \chi_k  $ and $\boldsymbol{\delta} $ are decision variables. Constraint (\ref{Eq:SG_opt_cons.}) determines the lower and upper bound on the electricity price.

In practice, this scheme is not appropriate for data centers, given the risk of performance degradation \cite{Wierman2014ICGCC}. Also, due to selfishness, a CP accepts the grouping proposal of SG if this partitioning maximizes its payoff. Otherwise, the CPs deviate and prefer to form other coalitions to increase their utilities.
Hence, the CPs prefer to distributively decide about the coalition formation in such a way to maximize their utilities, which does not necessarily lead to maximum profit for the SG. We will explain this process in the next section. The centralized scheme, in which the CPs are not selfish, can be used as a benchmark as it gives an upper bound on the payoff that can be obtained by the smart grid.
\subsection{Interactive Cooperative Game}
\label{SubSec:ICG}
In this section, we introduce the CPs' coalition formation mechanism and find stable coalitions.  Then, formulate the smart grid optimization in the interactive game model.
\subsubsection{Coalition Formation Mechanism}
\label{subsubsec:Coalition_Formation_Mechanism}
Having defined the characteristics of the CPs coalitional game
as well as the payoff sharing rule in Section \ref{SubSec:Coalitional_Game_Model}, we need to determine the cooperative strategies of the CPs. These strategies specify how the CPs can switch between the different coalitions.
 We define two strategies for any CP in the coalitional game: 1) Split from a coalition; 2) Merge into a coalition \cite{Niyato2011IEEE}. 

Consider a coalition $ \mathcal{S} $ formed between some CPs.
The CPs decide to \textit{split} from current coalition $ \mathcal{S} $  and formed multiple new coalitions $ \mathcal{S}^{'}$, where $ \mathcal{S}= \cup \mathcal{S}^{'}$, if the payoff obtained by all CPs in $ \mathcal{S}^{'} $ are larger or equal to their payoff in $ \mathcal{S} $, and at least the profit of one CP strictly improves, as follows:  
\begin{equation}
\label{Eq:Split}
\psi_j(\mathcal{S})\geq \psi_j(\mathcal{S}^{'}),\forall j \in \mathcal{S},
\textrm{and}~\exists k\in \mathcal{S},\psi_k(\mathcal{S})> \psi_k(\mathcal{S}^{'})
 \end{equation}

On the other hand, the CPs who are in multiple coalitions $ \mathcal{S} $ decide to merge into a single new coalition $ \mathcal{S}^{"} $, where $ \mathcal{S}^{"}= \cup \mathcal{S} $, if all CPs in all coalitions obtain larger or equal payoff in this new coalition than the former coalitions, and the profit of at least one CP strictly improves, i.e. 
\begin{equation}
\label{Eq:Merge}
\psi_j(\mathcal{S}^{"})\geq \psi_j(\mathcal{S}),\forall j \in \mathcal{S},
\textrm{and}~\exists k\in \mathcal{S},\psi_k(\mathcal{S}^{"})> \psi_k(\mathcal{S})
 \end{equation}
 
The time complexity of the coalition formation mechanism is determined by the number of merge and split operations and the size of the sub-partitions. In the worst-case scenario, each coalition needs to make a merge attempt with all the other coalitions. In the initial coalition structure in which each one of the $ N $ individual CPs is a federation, the first merge needs $ \frac{N(N-1)}{2} $ attempts in the worst-case. The second merge requires $ \frac{(N-1)(N-2)}{2} $ attempts and so on. As a result, the total worst-case number of merge operations is in $ O(N^3) $.
 In the worst-case, splitting a federation $ \mathcal{S} $ is in $ O(2^{|\mathcal{S}|}) $, which involves finding all the possible partitions of size two of the participating federations \cite{Mashayekhy2014TCC}.
\subsubsection{Stable Coalitions}
\label{Subsec:Stable_Coalitions}
After defining the cooperative strategies of the CPs, now, we should analyze the dynamic of the coalition formation process to find under what conditions it settles down to the stable coalitional states. We apply a Markov decision process (MDP) to obtain the solution.

Considering $\chi_k $ as a coalition structure, we define the state space of the CPs coalition formation as follows:
\begin{equation}
\label{Eq:Coalition_State}
\varOmega=\{\chi_{k}|k=\{1,..,B_N\}\}.
\end{equation}

The coalitional state will be changed if the strategies of some CPs change due to their decision for merging into or splitting from their current coalitions depend on the obtained payoff from different coalitions. Given the electricity prices announced by the smart grid, we denote the transition probability matrix for a coalitional state as follows:
\begin{equation}
\label{Eq:Prob_mtx}
\boldsymbol{T}(\boldsymbol{\delta})=
\left [
\begin{matrix}
  \eta_{1,1}(\boldsymbol{\delta}) & \eta_{1,k^{'}}(\boldsymbol{\delta})& ...& \eta_{1,B_N}(\boldsymbol{\delta}) \\
  . & . & .&.\\
  . & . & .&.\\
 \eta_{k,1}(\boldsymbol{\delta}) & \eta_{k,k^{'}}(\boldsymbol{\delta})& ...& \eta_{k,B_N}(\delta) \\
 . & . & .&.\\
 . & . & .&.\\
 \eta_{B_N,1}(\boldsymbol{\delta}) & \eta_{B_N,k^{'}}(\boldsymbol{\delta})& ...& \eta_{B_N,B_N}(\boldsymbol{\delta}) \\
\end{matrix}
\right ],
\end{equation}
where $ \eta_{k,k^{'}}(\boldsymbol{\delta}) $ is the probability that a coalitional state changes from $ \chi_{k} $ to $ \chi_{k^{'}} $. Suppose that $ \Bbbk_{k,k^{'}} \subseteq \mathcal{N} $ is the set of the CPs who decide to perform merge and split actions, which results in a change in the coalition state from $\chi_{k} $ to $ \chi_{k^{'}} $, where $ \chi_{k^{'}} $ is reachable from $ \chi_{k} $, given the strategies of the CPs in $ \Bbbk_{k,k^{'}}$. Then, $ \eta_{k,k^{'}}(\boldsymbol{\delta}) $ can be given by \cite{Arnold2002Elsevier, Niyato2010GLOBECOM}: 
\begin{equation}
\label{Eq:Transition_Probability}
\eta_{k,k^{'}}(\boldsymbol{\delta})=\prod_{~i \in \Bbbk_{k,k^{'}}} 
\sigma \tau_i(\chi_{k^{'}}|\chi_{k}) (1-\sigma) ^{N-|\Bbbk_{k,k^{'}}|},
\end{equation}
where $ \sigma $ denotes the probability of performing a merge and split strategy by a CP. Note that if $ \chi_{k^{'}} $ is not reachable from $ \chi_{k} $, $ \eta_{k,k^{'}}(\boldsymbol{\delta}) $ will be equal to zero.  $ \tau_i(\chi_{k^{'}}|\chi_{k}) $ is a best-reply rule and is defined as the probability of changing the coalitional state from $\chi_{k} $ to $ \chi_{k^{'}} $ due to the change in the strategy of CP $ i $. This probability can be calculated as follows:
\begin{equation}
\label{Eq:tau}
\tau_i(\chi_{k^{'}}|\chi_{k})=
\begin{cases}
    \varrho, & if ~\psi_i(\mathcal{S} \in \chi_{k^{'}} )\geq \psi_i(\mathcal{S}\in \chi_{k} ), \\
    \epsilon, & otherwise. 
\end{cases}
\end{equation}
for $ i \in \mathcal{S} \in \chi_{k} $, $ i \in \mathcal{S} \in \chi_{k^{'}} $, and $ \chi_{k}\neq \chi_{k^{'}} $. $ \varrho  \in (0,1]$ is a constant, and $ \epsilon $ is the probability that the CP acting irrationally (e.g. due to the lack of information), leads to obtain less profit and generally has a small value. 
The diagonal elements are calculated as:
\begin{equation}
\label{Eq:Transition_Probability_Digonal}
\eta_{k,k}(\boldsymbol{\delta})=1-\left(      \sum_{ \chi_{k^{'}} \in \varOmega \backslash \{\chi_k\} }     \eta_{k,k^{'}}(\boldsymbol{\delta})     \right) .
\end{equation}

Given the fixed pricing parameter $ \boldsymbol{\delta} $ determined by the SG, we should derive the stationary probability of the Markov chain
defined by the state space in (\ref{Eq:Coalition_State}) and the transition probability
in (\ref{Eq:Prob_mtx}). We denote the vector of the stationary probabilities by
$ \overrightarrow{\boldsymbol{p}}_\delta= [p_\delta(\chi_{1}), . . . , p_\delta(\chi_{k}), . . . , p_\delta(\chi_{B_N})]^T $.
It can be obtained by solving the following set of equations: 
$ \overrightarrow{\boldsymbol{p}}_\delta^{T}= \overrightarrow{\boldsymbol{p}}_\delta^{T} \boldsymbol{T}(\boldsymbol{\delta})$, 
and $ \overrightarrow{\boldsymbol{p}}_\delta ^{T}\overrightarrow{\boldsymbol{1}}=1$.
If the probability of taking the irrational decision approaches zero (i.e., $ \epsilon \longrightarrow 0^{+} $), there could be an ergodic set $  \mathbb{E}_c \subseteq \varOmega $ of states $ \chi_{k} $. Once all players enter a coalitional state in an ergodic set, they will remain in that set forever after. Ergodic set defines as follows.
\begin{mydef}
A set $ \mathbb{E}_c $ is ergodic if $ \eta_{k,k^{'}}(\boldsymbol{\delta}) = 0 $, for any $ \chi_{k}\in  \mathbb{E}_c $ and $ \chi_{k^{'}}\notin  \mathbb{E}_c $, and no nonempty proper subset of $  \mathbb{E}_c $ has this property.
 The singleton ergodic sets are named absorbing states. In other words,$  \chi_{k} $ is absorbing set, if $ \eta_{k,k}(\boldsymbol{\delta}) = 1 $ \cite{Arnold2002Elsevier}.
\end{mydef}
 The absorbing states are stable coalitional states. Reaching to an absorbing state, which is a stable coalitional state, no player has an incentive to change its strategy, given the prevailing coalitional state.
As a result, by finding the stationary probability of the Markov chain, we can derive the stable coalition states.
 \subsubsection{Optimization Formulation for the Smart Grid}	 	 
  \label{SubsubSec:OPt_SG2}
  For any given coalitional structure, the smart grid must decide on its optimal pricing parameters' vector $\boldsymbol{\delta } $. Hence, given the underlying coalitional game between the CPs, the optimization formulation for the smart grid, is a 4-tuple $ (\varOmega,\Xi,\eta_{k,k^{'}}(\boldsymbol{\delta}),U^{SG}(\chi_{k},\boldsymbol{\delta})) $, where $ \varOmega $ is a finite set of coalitional states as defined in (\ref{Eq:Coalition_State}), $ \Xi  $ is a finite set of actions $ \boldsymbol{\delta}$, $ \eta_{k,k^{'}}(\boldsymbol{\delta}) $ is the transition probability defined in (\ref{Eq:Transition_Probability}), and
 $ U^{SG}(\chi_{k},\boldsymbol{\delta}) $ is the immediate profit (utility) the SG obtained at coalitional structures $ \chi_{k} $ with action $ \boldsymbol{\delta}$ defined in (\ref{Eq:SG_Utility}).

The SG aims at optimizing its long-term profit through choosing the optimal policy. A policy is a mapping from a coalitional state $ \chi_{k}\in \varOmega $ to an action $\boldsymbol{\delta} \in \Xi $. 
The SG should determine its action in such a way satisfies the constraint on the electricity price in (\ref{Eq:SG_opt_cons.}).
Therefore, the optimization problem for the smart grid can be formulated as: 
\begingroup
\allowdisplaybreaks
\begin{align}
\label{Eq:SG_Utility_Op}
& \max_{\pi}   \qquad  \varUpsilon_{U,\pi}   \\
\textrm{s.t.} \qquad 
& \theta_{l} \leq \varUpsilon_{\theta,\pi}  \leq \theta_{h}, \forall j \in \mathcal{N},  \label{Eq:SG_Utility_Op_c1}
\end{align}
\endgroup
where:
  \begin{equation}
   \label{Eq:Expected_SG_Utility}
  \varUpsilon_{U,\pi} =\lim_{T\longrightarrow \infty} \inf \dfrac{1}{T}
  \sum_{t=1}^{T} E_\pi [U^{SG}(\boldsymbol{\delta}(t),\chi_{k}(t))],
  \end{equation}
   \begin{equation}
   \label{Eq:Expected_Price}
   \varUpsilon_{\theta,\pi} =\lim_{T\longrightarrow \infty} \sup \dfrac{1}{T}
     \sum_{t=1}^{T} E_\pi [{\theta_i}(\boldsymbol{\delta}(t),\chi_{k}(t))],
   \end{equation}
  where $ \chi_{k}(t) $ and $ \boldsymbol{\delta}(t) $ are the coalition structure and the vector of the pricing parameters at time t, respectively. $ E_\pi(.) $ denotes  the expectation over policy $ \pi $. 
  Policy $ \pi $ is defined as $ \pi = \{\varphi(\chi_{k},\delta )| k= \{1,..., B_N\},\forall\boldsymbol{\delta}\in\Xi\}$, where $ \varphi(\chi_{k}, \boldsymbol{\delta}) $ is the probability of taking action $\boldsymbol{\delta}$ at coalitional state $ \chi_{k} $.
          
To obtain the optimal policy of MDP as defined in (\ref{Eq:SG_Utility_Op})-(\ref{Eq:SG_Utility_Op_c1}), an equivalent linear programming (LP) model can be formulated. We use this equivalent model to find $ \phi(\chi_{k},\boldsymbol{\delta}) $, which is the stationary probability of the coalitional state and action. The LP model is defined as follows:
\begin{equation}
\label{Eq:LP_opt}
\max_{\phi(\chi_{k},\boldsymbol{\delta})} 
\sum_{{\chi_k} \in \varOmega} 
\left(  
\sum_{\boldsymbol{\delta} \in \Xi} U^{SG}(\chi_{k},\boldsymbol{\delta}) \phi(\chi_{k},\boldsymbol{\delta})
\right)
\end{equation}
s.t.
\begingroup
\allowdisplaybreaks
\begin{align}
&\theta_{l} \leq \sum_{\chi_{k} \in \varOmega} 
\left(  \sum_{\boldsymbol{\delta}\in \Xi} \theta_i(\chi_{k},\boldsymbol{\delta}) \phi(\chi_{k},\boldsymbol{\delta}) 
\right)\leq \theta_{h}, \forall i \in \mathcal{N}, \label{Eq:LP_C1}\\
& \sum_{\boldsymbol{\delta}\in \Xi} \phi(\chi_{k^{'}},\boldsymbol{\delta})= \sum_{\chi_{k} \in \varOmega}\left( 
\sum_{\boldsymbol{\delta}\in \Xi}  \phi(\chi_{k},\boldsymbol{\delta})\eta_{k,k^{'}}(\boldsymbol{\delta})\right), \label{Eq:LP_C3}\\
& \sum_{\chi_{k} \in \varOmega} \left(  \sum_{\boldsymbol{\delta} \in \Xi} \phi(\chi_{k},\boldsymbol{\delta}) \right)=1,\label{Eq:LP_C4}\\
& \phi(\chi_{k},\boldsymbol{\delta}) \geq 0, \forall \chi_{k} \in \varOmega, \forall \boldsymbol{\delta} \in \Xi, \label{Eq:LP_C5}
\end{align}
\endgroup
for $ \chi_{k^{'}} \in \varOmega $.

Given the optimal solution $ \phi^{*}(\chi_{k} ,\boldsymbol{\delta})$ of the LP model,  we can calculate the probability that the smart grid selects $ \boldsymbol{\delta} $ when the coalitional state of CPs is $ \chi_{k}  $. We denote this probability by $ \varphi^{*}(\chi_{k},\boldsymbol{\delta} ) $, and  calculate as follows:
\begin{equation}
\label{Eq:Opt_policy}
\varphi^{*}(\chi_{k},\boldsymbol{\delta} )=\dfrac{\phi^{*}(\chi_{k} ,\boldsymbol{\delta})}{ \sum_{\boldsymbol{\delta}^{'} \in \Xi} \phi^{*}(\chi_{k} ,\boldsymbol{\delta}^{'})} ,
\end{equation}
for $ \sum_{\boldsymbol{\delta}^{'} \in \Xi} \phi^{*}(\chi_{k} ,\boldsymbol{\delta}^{'})> 0 $. 
Given the values of  $ \varphi^{*}(\chi_{k},\boldsymbol{\delta} ) $, the optimal policy is then defined as $ \pi^{*}=\{\varphi^{*}(\chi_{k},\boldsymbol{\delta} )|k= \{1,..., B_N\},\forall \boldsymbol{\delta}\in \Xi\}$

As we mentioned in section \ref{Subsec:Stable_Coalitions}, depending on the optimal policy of the smart grid, the stationary probability of the coalitional state can be derived. For an optimal policy $ \pi^{*} $, we denote the stationary probability of coalitional state $ \chi_{k} $ by $ \pi^{*}(\chi_{k}) $. 
The vector of the stationary probabilities is denoted by
$ \overrightarrow{\boldsymbol{p}}_{\pi^{*}}=
 [p_\pi^{*}(\chi_{1}), . . . ,p_\pi^{*}(\chi_{k}), . . . , p_\pi^{*}(\chi_{B_N})]^T $,  
and can be computed by solving the following equations:
$ \overrightarrow{\boldsymbol{p}}_{\pi^{*}}^{T}= \overrightarrow{\boldsymbol{p}}_{\pi^{*}} ^{T}\boldsymbol{T}_{\pi^{*}}$, 
and $ \overrightarrow{\boldsymbol{p}}_{\pi^{*}}\overrightarrow{\boldsymbol{1}}=1$.
In this regard, we compute the average profit of the smart grid as:
\begin{equation}
\label{Eq:Avg_rev_SG}
\overline{U^{SG}}=\sum_{\chi_{k} \in \varOmega} 
p_\pi^{*}(\chi_{k})
\left(  
\sum_{\boldsymbol{\delta}\in \Xi} \varphi^{*}(\chi_{k},\boldsymbol{\delta} )U^{SG}(\chi_{k},\boldsymbol{\delta})
\right),
\end{equation}
and, the average cost of content provider $ i $ is obtained from:
\begin{equation}
\label{Eq:Avg_rev_CP}
\overline{ \psi_i}=
\sum_{\chi_{k} \in \varOmega} 
p_\pi^{*}(\chi_{k})
\left(  
\sum_{\boldsymbol{\delta} \in \Xi} \varphi^{*}(\chi_{k},\boldsymbol{\delta} )  \psi_i(\chi_{k},\boldsymbol{\delta})
\right).
\end{equation}

Again, if the probability of irrational decision approaches zero, there could be an ergodic set $  \mathbb{E}_c \subseteq \varOmega $ of absorbing states. 
\section{Experimental Results}
\label{Sec:Results}
\subsection{ Parameter Setting}
\label{Subsec:param_setting}
We consider six DCs, which are operated by independent CPs, that are geographically located in different regions in the United States \cite{Zhou2015}.
The configuration parameters of DCs include the number of hosts ($ M$) and the value of parameter $ a$, the idle and peak power of the hosts, as well as PUE ($ \gamma$) of each DC are reported in Table \ref{Table:SimParameters}. Although the value of PUEs in our model can be constant or variable, here, for simplicity we assume constant PUE values for DCs under all loads. However, in practice, it varies with the IT load and environmental conditions and improved at lower loads. As a result, when some workloads migrated from a DC connected to a bus with a high electricity price to the other one with a lower price, we could gain more than we are currently earning by assuming the constant PUEs.

The cost of a VM migration from CP $ i $ to CP $ j $ is calculated as the product of the cost rate of transferring data, data rate, and the time of migration. Data cost rate is set to $ 0.001$ \$/GB according to the Amazon  EC2 data transfer pricing \cite{Amazon}, and we assume that data would be sent at a fixed rate of $ 100 $ Mbit/s. Migration time is a random number from the Normal distribution with mean $ 554 $ sec. and the standard deviation of $ 364 $ sec. \cite{Guazzon2014CCGrid}. 
   \begin{table}[t]
  \centering
  	\rowcolors{2}{gray!20}{white}
  \caption{Simulation parameters  \cite{Guazzon2014CCGrid}.} 
  \label{Table:SimParameters}
  \begin{small}
  \begin{tabular}{cccccc}  
  \hline
  $DC \# $  & $ M$ & $ a$ & PUE & $ P^\textrm{idle} (KW)$&  $  P^\textrm{peak}$  $(KW)$ 
   \\ \hline \hline
  $1 $ & $ 15000 $ & $ 1 $ & 1.3& $ 0.086  $ & $ 0.274 $\\ 
  $2 $  & $12000$ &$ 2 $ & 1.5& $ 0.143  $& $ 0.518 $  \\ 
  $3 $   & $ 10000$ & $ 3 $ & 1.3& $ 0.490 $& $ 1.117 $\\ 
  $4 $ & $ 20000$ & $ 1 $ & 1.6& $ 0.086 $& $ 0.274  $\\ 
  $5 $   & $15000$ & $ 2 $& 1.8& $ 0.143  $& $ 0.518 $ \\ 
  $6 $ & $10000 $ & $ 3 $ & 1.1  & $ 0.490 $& $ 1.117 $\\
   \hline
  \end{tabular} 
    \end{small}
  \end{table}

 The dynamic computing requests are simulated using the one-day HP request trace reported in \cite{Zhou2015}. It is scaled proportionally to the total capacity of data centers. Then, the workload of each hour randomly is distributed among data centers based on their capacity.
We took hourly demands of six locations on 27th August 2018 as the background power load for calculating $ G_i $s in (\ref{Eq:SG_Utility}) \cite{nyiso,neiso,energyonline}.
The lower and upper boundaries of electricity prices are set to be 8  and 25 cents/kWh based on the data from \cite{EIA}.
The parameters of the interactive coalitional game are $ \sigma=0.5 $, $ \varrho=0.99 $, $ \epsilon=0.01 $, $ \alpha_1=0.3 $, and $ \alpha_2=0.7 $ \cite{Niyato2011IEEE}. 
We use MATLAB for simulation and YALMIP toolbox \cite{lofberg2004yalmip} along with CPLEX \cite{cplex} solver for solving the optimization problems.
  \begin{figure*}[t]
  \centering
  \includegraphics[width=0.95\textwidth]{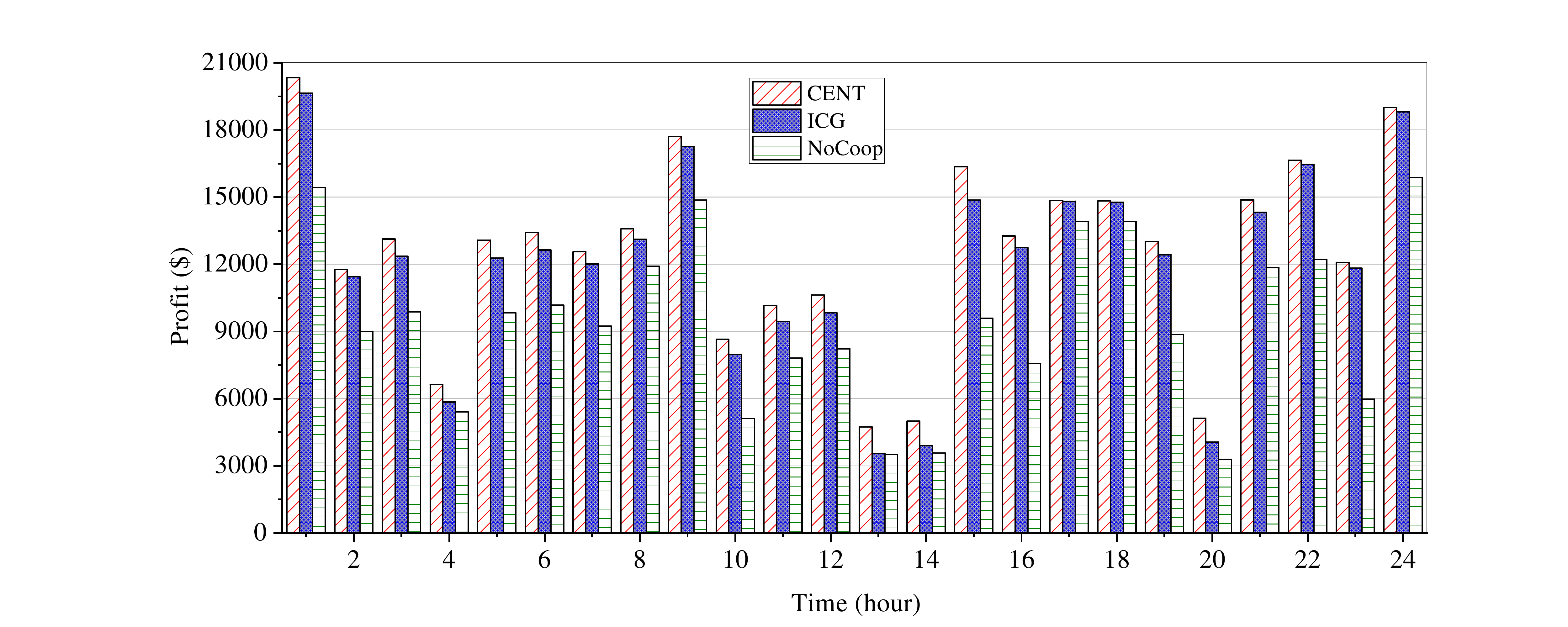}
  \caption{The SG's profit for three schemes in the scheduling period.}
  \label{Fig:Results_USG_Profit}
  \end{figure*}
  \begin{figure*}[t]
  \centering
  \includegraphics[width=0.95\textwidth]{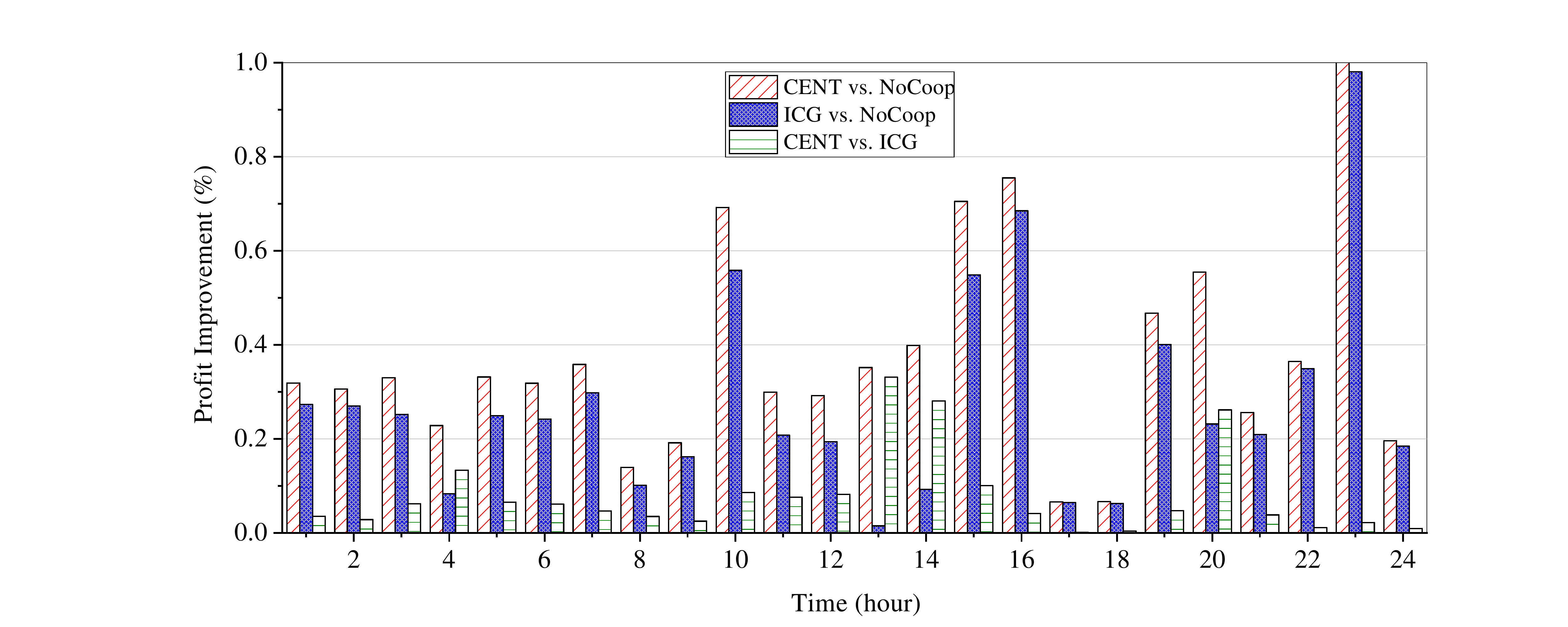}
    \caption{Profit improvement of SG in the scheduling period.}
  \label{Fig:Results_USG_Profit_Imp}
  \end{figure*}
    \begin{figure*}[t]
    \centering
    \includegraphics[width=0.92\linewidth]{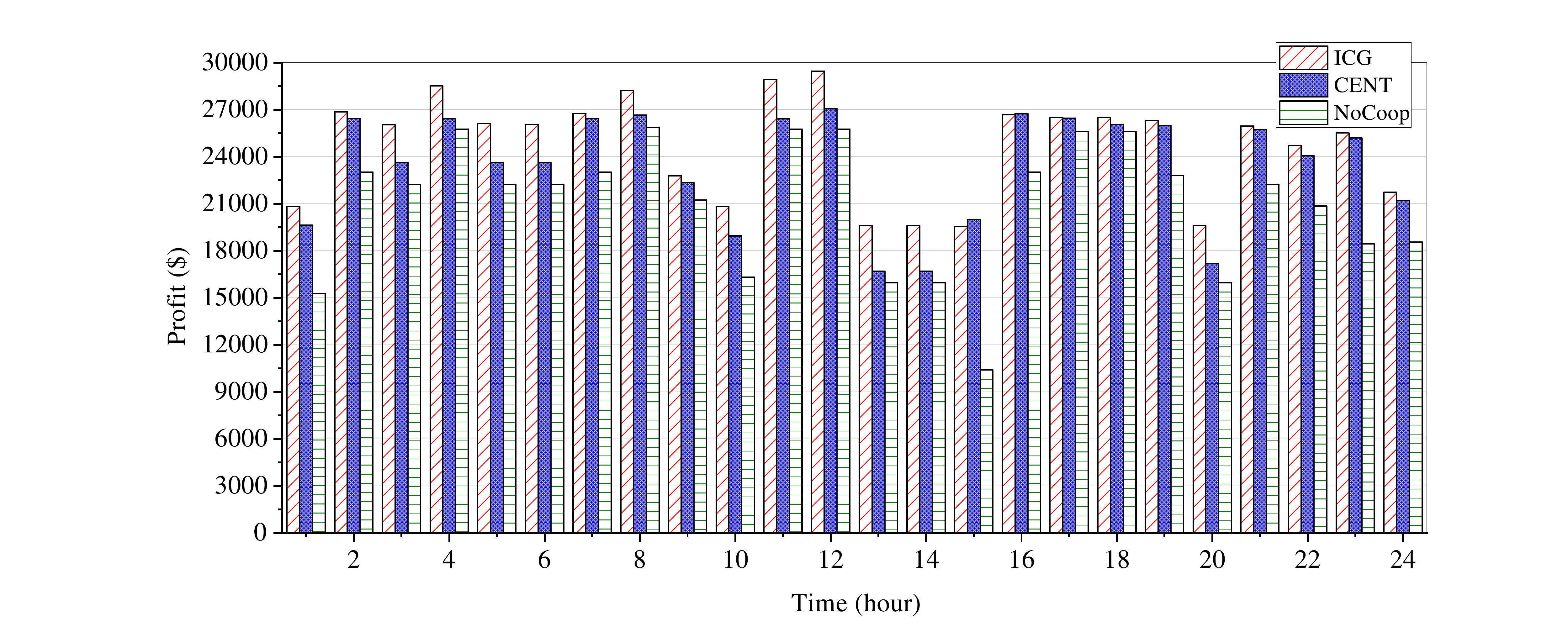}
   \caption{The aggregated profit of CPs for three schemes in the scheduling period.}
    \label{Fig:Results_UCP_Profit}
    \end{figure*}
\subsection{Simulation Results and Evaluation}
\label{Subsec:sim_results}
In addition to the proposed interactive cooperative game model (\textit{ICG}), we also simulate two other approaches: a centralized approach introduced in Section \ref{SubSec:SG_Utility_CentScheme} (we call it \textit{CENT}), and a noncooperative scheme (we named it \textit{NoCoop}). 

In the CENT approach, we assume that the smart grid decides about the CPs coalition partition, the CPs are not selfish, and they follow the SG suggestion to form their coalitions. 
In NoCoop, the cloud providers individually process their workload without forming any coalition. Hence, regardless of the selected strategy by the SG ($ \boldsymbol{\delta} $ in the pricing function), CPs have the specific power demand. Therefore, the value of $ e_j $s will be fixed.
Consequently, the second term in  (\ref{Eq:SG_Utility}), has a specific fixed value. 
But, depending on the smart grid strategy, the unit electricity price will be changed, resulting in different values for the first term in  (\ref{Eq:SG_Utility}).
Since the SG is a rational player, chooses the minimum value for $ \boldsymbol{\delta}  $ to maximize the electricity price, consequently, its profit.
Therefore, in the simulation of the noncooperative scheme, we report the results for the strategy with the maximum SG's profit. 

In what follows, we evaluate the effectiveness of our model in terms of average profits obtained by the SG operator as well as by the cloud providers.
\subsubsection{SG Profit}
\label{Subsec:Results_SG_profit}
Figure \ref{Fig:Results_USG_Profit} shows the resulting SG’ profit during the scheduling period.
We also show the profit improvement of ICG and CENT compared to the NoCoop scheme, as well as the profit improvement percentage earned through CENT in comparison to ICG in Figure \ref{Fig:Results_USG_Profit_Imp}.

As can be seen from Figures \ref{Fig:Results_USG_Profit} and \ref{Fig:Results_USG_Profit_Imp}, the two cooperative approaches (i.e., ICG and CENT) improve substantially the total SG' profit compared to noncooperative scheme, which demonstrates the positive effect of CPs cooperation on the SG profit when the SG chooses its strategy by considering this cooperation.  Based on these figures, the average one-day's profit of the SG in CENT, ICG and NoCoop is equal to about 125560, 11930, and 9456, respectively.
Indeed, the average profit improvement in CENT and ICG compared to the NoCoop case is 38 and 28\% on average, respectively. As we mentioned in Section \ref{SubSec:SG_Utility_CentScheme}, CENT approach as a benchmark gives an upper bound on the profit can be earned by SG through the CPs cooperation, and Figure \ref{Fig:Results_USG_Profit_Imp} shows that this bound is not much larger than the value obtained by ICG approach (8\% on average). Nonetheless, the SG operator can design mechanisms to encourage CPs to follow its suggestion to achieve the maximum payoff.
   \begin{figure*}[t]
    \centering
   \begin{subfigure}[t]{\columnwidth}
  \centering
  \includegraphics[width=0.88\linewidth]{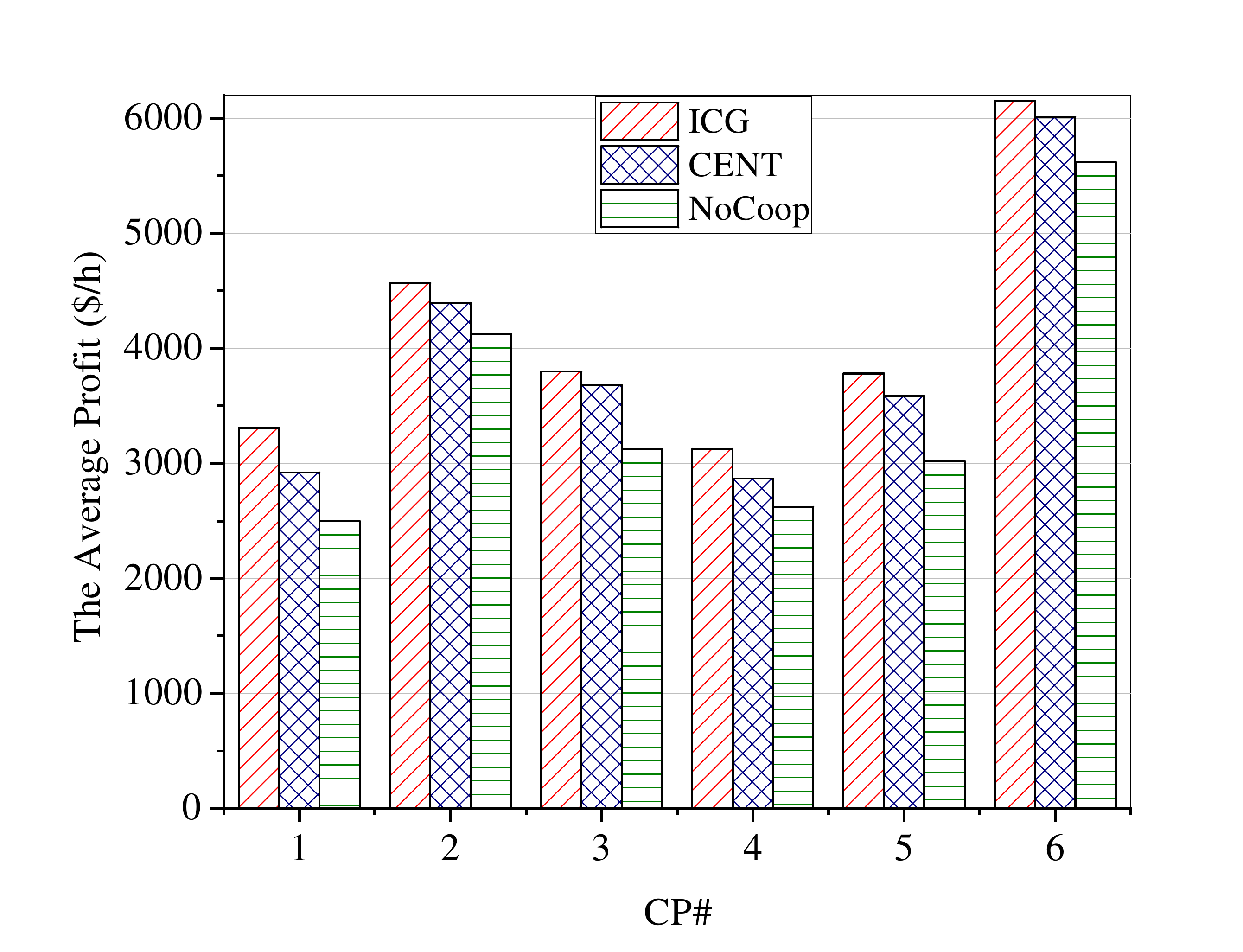}
   \caption{}
  \end{subfigure}
   \begin{subfigure}[t]{\columnwidth}
  \centering
  \includegraphics[width=0.88\linewidth]{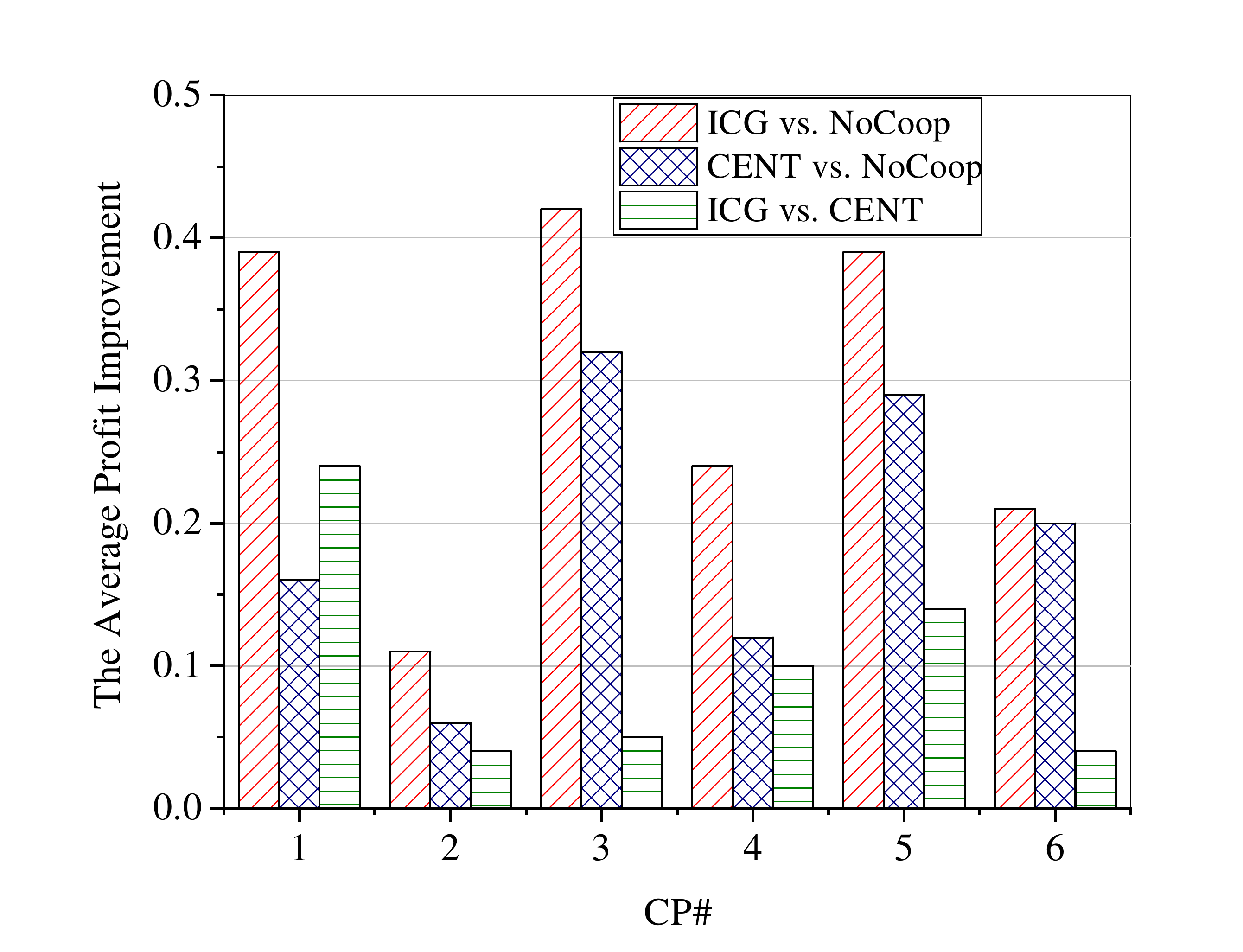}
   \caption{}
  \end{subfigure}
  \caption{Simulation results for clouds' profit (a) The average profit of CPs for three schemes (b) The average profit improvement of CPs for ICG and CENT compared to NoCoop, as well as CENT in comparison to ICG.}
  \label{Fig:Results_UCP_Profit_Imp}
   \end{figure*}
\subsubsection{CPs Profit}
\label{Subsec:Results_CP_profit}
Figure \ref{Fig:Results_UCP_Profit} shows the aggregated profit earned by the CPs during different hours, and Figure \ref{Fig:Results_UCP_Profit_Imp} presents the average profit and the profit improvement for different CPs in ICG and CENT compared to NoCoop, as well as the profit improvement of ICG vs. CENT approach. From these figures, we observe that cooperative participation of CPs in demand response programs yields more profit compared to the noncooperative scheme. This improvement is 29\%, and 19\% for ICG and CENT, on average. Nonetheless, the profit improvement of  CPs are not the same, and they earn different profits depending on their configurations and contributions.

However, the aggregated profit of  ICG is larger than the CENT approach during most hours, the difference in the average profit is insignificant (The ICG on average leads to 10\% more benefit than the CENT model). But, as the CPs are rational, selfish entities, they prefer to distributively decide on the cooperation process to achieve the maximum payoff, unless the SG applies the other mechanism to mitigate this difference.
In summary and based on the simulation results, it is clear that CPs can maximize their payoffs in DR programs by cooperative workload management, and the SG can also gain from this cooperation by right setting its strategy.
\section{Conclusion}
\label{Sec:Conclusion}
In this paper, we have focused on the interaction of the smart grid, and multiple cooperative cloud providers in the context of a dynamic electricity pricing scheme.
We aimed to formulate the process of dynamic electricity pricing
by the SG along with the cooperative workload management
of independent CPs to analyze the impact of decisions taken by each side on the other.
We have modeled this scenario as a two-stage Stackelberg game interleaved with a coalitional game, we named interactive cooperative game (ICG). In this game, the SG as the leader specifies the pricing functions, and the CPs as followers manage their workload collaboratively so that minimize their cost.
We have applied a coalitional game formulation for analyzing the cooperation process among CPs, and an optimization model based on the constrained Markov decision process for finding the optimal policy by the SG.

We have performed thorough simulations to evaluate the proposed scheme.
Simulation results showed that ICG significantly improves the profit of SG as well as the CPs compared to the noncooperative case. The values of SG's utility improvement for ICG approach is 28\%, on average,  compared to the noncooperative scheme. The CPs profits are also increased by 29\%, on average, in ICG scheme.
\balance
\bibliographystyle{IEEEtran}
\bibliography{IEEEabrv,MyReferences}
\newpage
 \onecolumn
 \begin{landscape}
 \pagestyle{empty}
 \small
 \rowcolors{1}{gray!20}{white}
\renewcommand*{\arraystretch}{1.4}
\begin{longtabu} to \columnwidth {>{\small}X[1,l,m] >{\small}X[0.6,l,m]  >{\small}X[0.7,l,m]  >{\small}X[0.6,l,m] >{\small}X[0.5,l,m]  >{\small}X[1,l,m] >{\small}X[0.1,l,m] } 
   \hiderowcolors
    \caption{ Summery of related works  on DCs participating in DR programs.}
    \tabularnewline
      \hline
    Topic&DR Scenario& DC Operator Objective&  SG Objective& 
    Problem Formulation&   Other Features&   Ref.
     \tabularnewline
     \hline
     \endfirsthead   
     \caption{ Summery of related works  on DCs participating in DR programs (Cont.)}
     \tabularnewline        \hline      
   Topic&  DR Scenario& Objectives of CPs&  Objectives of SG& 
    Problem Formulation&   Other Features&  Ref.
      \tabularnewline \hline
       \endhead
    \showrowcolors
  Service request routing in CC jointly with power flow analysis in SG&
Regulated and deregulated market&
NA&
Load balancing and robustness&
Optimization&
NA&
\cite{Mohsenian-Rad2010SmartGridComm}  
  \tabularnewline 
Find best location for building a DC&
Regulated and deregulated market  &
Minimizing 1. carbon footprint, 2. energy \& bandwidth \& carbon tax cost,  2. average service latency&
 NA&
MILP&
 NA&   
\cite{Mohsenian-Rad2010}    
 \tabularnewline  
 Minimizing the total electricity cost of IDCs while guaranteeing QoS&
Multi-electricity-market&
Minimizing total electricity cost &
NA &
 Constrained MILP&
 Consider delay and workload constraints&
 \cite{Rao2010IEEE, Rao2012}
\tabularnewline   
\strut
Model the market power of IDC and formulate electricity cost minimization &
NA&
Minimizing  electricity cost &
NA &
MILNP&
Consider DCs as price makers, and take into account delay and workload constraints&
\cite{Wang2011GLOBECOM,Wang2012SG}
\tabularnewline   
\strut
Minimize the IDC operation risk &
Electricity spot markets&
Minimizing operation risk &
NA&
Bilevel programming&
Consider correlation between workload and spot electricity price&
 \cite{Rao2011SmartGrid, yu2012IEEE}    
  \tabularnewline     
Adjust DCs load to balance the unstable solar input&
NA&
NA&
Minimizing power loss&
Genetic algorithm&
Analyzed the impact of flexibility and delay of DC actions \& Predict renewable generation in \cite{luo2019}&
\cite{luo2018ACM, luo2019}
 \tabularnewline     
 \strut
 Dynamic load distribution and migration policies&
Peak Power&
Minimizing electricity cost&
NA&
NA&
Consider transient cooling effect &
\cite{le2011ACM}
 \tabularnewline     
 \strut
Minimize electricity cost under diverse delay requirements &
Multi-electricity-market &
Minimizing energy cost&
NA&
CMIP&
NA&
\cite{fan2016CSCloud}
 \tabularnewline     
GLB scheme for distributed price-maker IDCs &
Deregulated power market&
Minimizing energy cost+revenue loss of degraded service performance&
NA&
Optimization&
Consider impact of GLB upon electricity prices&
\cite{yu2016TCC} 
\tabularnewline     
Joint inter- and intra-data center workload management&
NA&
Minimizing energy cost&
NA&
Constrained nonlinear optimization&
NA&
\cite{guo2014Elsevier}
\tabularnewline     
Dynamic pricing negotiation&
Dynamic pricing&
Minimizing energy cost&
NA&
Optimization&
NA&
\cite{Li2013ICDEW}
\tabularnewline     
Minimize energy cost of DCs&
Tiered pricing&
Minimizing energy cost&
NA&
Convex optimization&
DCs powered by power grid and local renewable energy generations&
\cite{Wang2016IGSEC}
\tabularnewline     
Optimize self-power consumption of a distributed cloud with on-site renewable generation&
NA&
Minimizing energy cost&
NA&
NA&
Virtually exchange renewable energy between DCs&
 \cite{camus2018CLUSTER}
 \tabularnewline     
 \strut
Minimize the operational cost of cloud for fair request allocations&
Deregulated market&
Minimizing the operational cost&
NA&
Optimization&
Propose a fast approximation algorithm for solving the problem&
 \cite{xu2013IEEE}
 \tabularnewline     
Cost-efficient workload scheduling&
NA&
Reducing the penalties cost of scheduling scheme&
Smooth the load variation&
Multi-objective optimization&
Consider batch and interactive workloads&
\cite{hu2016ICC,hu2017TSC}
\tabularnewline     
Optimal coordinated operation framework for minimizing the overall electricity cost&
Day-ahead and real-time wholesale market&
Minimizing the overall electricity cost&
NA&
NA&
Consider cost of purchasing power from main grid, operation cost of onsite generations, carbon emission cost, \& consider batch and interactive workloads&
\cite{qi2016toIEEE}
\tabularnewline     
Greening scheduling of cloud DCs in an economical way&
NA&
Minimizing 1.energy cost, 2. carbon emission&
NA&
Complex MILP&
Consider fluctuates in power prices, energy supply, and dynamics of request
 &
\cite{gu2018Elsevier}
\tabularnewline     
\strut
Manage the uncertainties in IDC operations under SG environment&
Forward and spot market&
Minimizing weighted sum of operation risk and expected energy cost&
 NA&
Two-stage stochastic programming&
Consider Self-generation DCs&
\cite{yu2013IEEE} 
\tabularnewline     
Electricity bill capping&
Locational pricing&
Minimizes the electricity cost, maximize throughput within cost budget &
NA&
MILP&
\begin{itemize}
\item Consider the energy consumption of cooling and network equipment
\item Consider the DCs as price maker
\end{itemize}&
\cite{zhang2012IEEE}
\tabularnewline     
DCs’ coupled decisions of choosing utility companies and scheduling workload&
Real-time pricing&
Minimizing the energy and risk cost&
Maximizing power selling revenue and minimizing peak-to-average ratio&
 Matching game with externalities&
Consider uncertainty in the renewable generation and workloads arrival&
\cite{bahrami2018SG}
\tabularnewline     
Study interactions between SG and DCs by considering the active decisions on both sides&
Regional dynamic pricing&
Minimizing energy cost&
Power load balancing&
Two-stage problem/ bilevel optimization&
\begin{itemize}
\item Consider the effect of load distribution on the SG
\item Study the impact of background load prediction error \cite{Wang2015IEEE} 
\end{itemize}
&
\cite{Wang2014, Wang2015IEEE} 
\tabularnewline     
Model interaction between SG, and cloud computing&
Two location-dependent dynamic  pricing&
Maximizing profit &
Maximizing profit and performing load balancing&
Two-stage sequential game&
Consider power flow analysis, and distributed renewable generation in the SG \cite{Wang2014IEEE}&
\cite{Wang2013ISGT,Wang2014IEEE}
\tabularnewline   
\strut
Model interactions among SG, cloud computing, and other load devices&
 Location-dependent real-time pricing&
Maximizing profit &
Maximizing profit and performing load balancing&
Two stage Stackelberg game+non-cooperative game&
NA&
\cite{Wang2014ISGT}
\tabularnewline   
 Study DR of geo-distributed DCs using electricity pricing signals&
 Real-time pricing&
Cost minimization &
 Profit maximization&
Two-stage Stackelberg +non-cooperative game &
\begin{itemize}
\item  Utilize workload shifting and dynamic server allocation
\item Consider dependency between utilities
\end{itemize} &
 \cite{Tran2014USENIX, Tran2015SmartGrid}
 \tabularnewline  
 Study bilateral electricity trade between SGs and DCs with hybrid green sources&
Real-time pricing&
Cost minimization &
Profit maximization&
 Stackelberg game+non-cooperative game&
DCs participate in net metering program to sold excess renewable energy  to SGs&
\cite{Zhou2015ACM,  zhou2016IEEE}
\tabularnewline   
Energy-aware resource allocation by using of renewable energy sources &
NA&
Sustain the energy of DCs using renewable energy sources (RES) &
Profit maximization&
Stackelberg games
Consider users' SLA and QoS&
\cite{aujla2017IEEECC}
\tabularnewline  
Online cost minimization of distributed DCs and EVs of their employees&
NA&
 Cost minimization problem of DCs and EVs of their employees&
NA&
Stochastic programming/ Lyapunov optimization&
\begin{itemize}
\item Consider uncertainties in DC workloads, electricity prices, and EV energy demands
\item Consider EV charging delays and peak power limits
\end{itemize}
&
 \cite{yu2016IoT} 
\tabularnewline  
Risk-aware optimization framework allows DCs  to operate as controllable load resources&
Real-time and emergency DR in wholesale market&
NA&
Profit maximization&
Optimization&
Consider uncertainties in workload migration time,  and obtained payoff from DR&
 \cite{Wang2013ICDCS}
\tabularnewline  
\strut
Efficient incentive mechanism to elicit DR from geo-distributed clouds&
incentive DR/ bidding&
Social welfare maximization (the aggregated utility of the cloud and  SGs&
Social welfare maximization&
Auction mechanism&
Strike a balance among the economic efficiency, truthfulness and the computational efficiency&
\cite{Zhou2015INFOCOM, zhou2018IEEE}
\tabularnewline 
DC cost reduction  by exploiting the diversity in the electricity price in multiple market&
Bidding in day-ahead and real-time market&
Profit maximization&
NA&
Stochastic optimization&
Consider users SLA, risk, and statistical characteristics of workload and electricity prices&
\cite{Ghamkhari2014INFOCOMWKSHPS}
\tabularnewline
\strut
Cost reduction via jointly optimizing electricity procurement and GLB&
Bidding in real-time and day-ahead markets&
Minimizing total electricity and bandwidth cost&
NA&
Auction&
Consider a distribution for electricity price and workload arrival&
\cite{zhang2017INFOCOM}
\tabularnewline
Study interactions of GLB and electricity supply chains&
Bidding in  day-ahead wholesale market&
Minimizing energy and GLB cost&
NA&
Auction&
NA&
\cite{Camacho2014ACM}
\tabularnewline
Optimal trading in the forward electricity market &
Forward market&
Minimizing cost of electricity, switching, response time, network delay,  carbon emission&
NA&
Convex optimization&
Consider electricity price variation, renewable availability, cooling  conditions, the effect of carbon emission and server switching&
\cite{paul2016Elsevier}
\tabularnewline  
\strut
Optimal DR capabilities of IDCs considering uncertainty in workload arrival&
Day-ahead market&
Minimizing expected electricity bill&
NA&
 MILP-based stochastic optimization&
Consider cost of electricity consumption, emission, migration, and 
revenue from DR provision&
\cite{chen2013TSG}
\tabularnewline  
\strut
Study Emergency DR of colocations by delaying the tenants’ batch workload and diesel usage&
Emergency DR&
Minimizing cost of tenants’ performance and operator’s diesel generator&
NA&
Auction&
Present an online algorithm by considering users' QoS in \cite{Sun2016ACM}&
\cite{Zhang2016TOMPECS, Sun2016ACM}
\tabularnewline 
\strut
Balance excess renewable energy in colocation DCs&
Discounted energy prices in form of energy credits&
Maximizing profit of cloud broker&
NA&
Optimization&
NA&
\cite{abada2018Springer}
\tabularnewline 
Enable reliable DR through DR capacity aggregation&
 Capacity bidding program&
  Maximizing the expected profit&
  NA&
  Coalitional game&
  Consider uncertainty in DR capacities&
  \cite{niu2016}
  \tabularnewline 
  \strut
  Energy cost reduction via aggregately bidding &
  Bidding in day-ahead forward market&
  Minimizing the electricity bill&
  NA&
  Coalitional game&
  Consider uncertainty in DCs' power demand&
  \cite{yu2017ACM} 
  \tabularnewline 
   \hiderowcolors
     \hline   
      \label{Table:RelatedWorks1}
  \end{longtabu}
 \end{landscape}
 \pagestyle{plain}
\pagestyle{empty}
\begin{landscape}
\small
\rowcolors{1}{gray!20}{white}
\renewcommand*{\arraystretch}{1.4}
 \begin{longtabu} to \columnwidth {>{\small}X[1,l,m] >{\small}X[0.8,l,m] >{\small}X[0.6,l,m]  >{\small}X[0.8,l,m] >{\small}X[0.1,l,m] } 
    \hiderowcolors
    \caption{ Summery of related works on energy management of cloud federations.}
\tabularnewline
  \hline
Topic& Objectives of CPs& Problem Formulation& Other Features&
Ref.
 \tabularnewline
 \hline
  \endfirsthead     
 \tabularnewline
\hline      
Topic& Objectives of CPs& Problem Formulation& Other Features&
Ref.
   \tabularnewline
   \endhead
    \showrowcolors
    Energy-aware (re)allocation of VMs&
    Minimizing energy consumption or carbon emission&
    Constraint Programming& 
    Consider the cloud users' SLA, and DC features&
    \cite{dupont2012e-Energy}
    \tabularnewline
      Reducing the carbon dioxide emissions in federated cloud
      ecosystems&
     Reducing carbon emission&
      NA&
      NA&
      \cite{giacobbe2015SustainIT}
     \tabularnewline
     \strut
     Load balancing among federated micro-datacenters powered by renewable energies&
      Minimizing non-renewable energy cost&
     Markov Chain/ Mean Field Analysis&
     NA&
     \cite{neglia2016ACM}
     \tabularnewline
     \strut
     Distributed  approach for federation formation&
     Maximizing individual profit of clouds&
     Cooperative game theory&
     NA&
     \cite{Guazzone2014CCGrid}
     \tabularnewline
     \strut
     Energy-aware resource and revenue sharing mechanism for cloud federations&
     Maximizing social welfare of CPs&
     Coalition game theory&
     Consider demand variations of internal users&
     \cite{Hassan2015IEEE}
     \tabularnewline
     Energy-aware scheduling policy for commercial and academic cloud federations&
     Minimizes the energy consumption of a federation while maintaining its
     performance&
     NA&
     NA&
     \cite{Kecskemeti2013Wiley}
     \tabularnewline
     VM allocation, cost management, and cooperation formation &
      Minimize the expected cumulative cost of all cooperative
      clouds&
       Coalitional game/ Network formation game/ Stochastic programming &
      Consider uncertainties of users’ demand, power price, and renewable power source&
      \cite{Kaewpuang2014IEEE}
      \tabularnewline
       DC energy flexibility exploitation for smart grid integration \&
       business scenarios for participating DC federations in DR
       &
      Adapt the DCs energy demand to improve the SG's stability &
      MINLP&
      Consider electrical and thermal energy, and workload relocation flexibility&
      \cite{cioara2019Elsevier}
      \tabularnewline
    \hiderowcolors
   \hline   
    \label{Table:RelatedWorks2}
 \end{longtabu}
 
\end{landscape}

\end{document}